\documentclass[12pt,preprint]{aastex}

\newcommand{\bdv}[1]{\mbox{\boldmath$#1$}}

\def\au{{\rm AU}}

\def\masyr{{\rm mas}\,{\rm yr}^{-1}}
\def\kpc{{\rm kpc}}
\def\mas{{\rm mas}}

\def\muas{\mu{\rm as}}

\def\max{{\rm max}}
\def\min{{\rm min}}
\def\rel{{\rm rel}}

\def\eff{{\rm eff}}

\def\e{{\rm E}}
\def\bpi{{\bdv\pi}}
\def\bmu{{\bdv\mu}}

\def\bgamma{{\bdv\gamma}}

\begin{document}
\title{OGLE-2017-BLG-1434Lb: Eighth $q<1\times 10^{-4}$ Mass-Ratio 
Microlens Planet Confirms Turnover in Planet Mass-Ratio Function}

\author{\textsc{
A. Udalski$^{1}$,
Y.-H. Ryu$^{2}$, 
S. Sajadian$^{3}$,
A. Gould$^{2,4,5}$, 
\and 
P. Mr\'{o}z$^{1}$, 
R. Poleski$^{1,5}$, 
M. K. Szyma\'{n}ski$^{1}$, 
J. Skowron$^{1}$, 
I. Soszy\'{n}ski$^{1}$, 
S. Koz{\l}owski$^{1}$,
P. Pietrukowicz$^{1}$, 
K. Ulaczyk$^{1}$,
M. Pawlak$^{1}$,
K. Rybicki$^{1}$,
P. Iwanek$^{1}$\\
(OGLE Collaboration)\\
M. D. Albrow$^{6}$,
S.-J. Chung$^{2,7}$, 
C. Han$^{8}$, 
K.-H. Hwang$^{2}$, 
Y. K. Jung$^{9}$,
I.-G. Shin$^{9}$, 
Y.~Shvartzvald$^{10,^{\dag}}$, 
J. C. Yee$^{9}$, 
W.~Zang$^{11,12}$,
W. Zhu$^{13}$, 
S.-M. Cha$^{2,14}$, 
D.-J. Kim$^{2}$, 
H.-W. Kim$^{2}$, 
S.-L. Kim$^{2,7}$, 
C.-U. Lee$^{2,7}$,
D.-J. Lee$^{2}$, 
Y. Lee$^{2,14}$, 
B.-G. Park$^{2,7}$,
R. W. Pogge$^{5}$ \\
(KMTNet Collaboration)\\
V. Bozza$^{15,16}$,
M. Dominik$^{17}$,
C. Helling$^{17}$,
M. Hundertmark$^{18}$,
U.G. J{\o}rgensen$^{19}$,
P. Longa-Pe{\~n}a$^{20}$,
S. Lowry$^{21}$,
M. Burgdorf$^{22}$,
J. Campbell-White$^{21}$,
S. Ciceri$^{23}$,
D. Evans$^{24}$,
R. Figuera Jaimes$^{17}$,
Y.I. Fujii$^{19,25}$,
L.K. Haikala$^{26}$,
T. Henning$^{7}$,
T.C. Hinse$^{2}$,
L. Mancini$^{7,27,28}$,
N. Peixinho$^{20,29}$,
S.Rahvar$^{30}$,
M. Rabus$^{31,7}$,
J. Skottfelt$^{32}$,
C. Snodgrass$^{33}$,
J. Southworth$^{24}$,
C. von Essen$^{34}$\\
(MiNDSTEp Collaboration)\\
} 
}

\affil{$^{1}$Warsaw University Observatory, Al. Ujazdowskie 4, 00-478 Warszawa, Poland}
\affil{$^{2}$Korea Astronomy and Space Science Institute, Daejon 34055, Korea}
\affil{$^{3}$Department of Physics, Isfahan University of Technology, Isfahan 84156-83111, Iran} 
\affil{$^{4}$Max-Planck-Institute for Astronomy, K\"{o}nigstuhl 17, 69117 Heidelberg, Germany}
\affil{$^{5}$Department of Astronomy, Ohio State University, 140 W. 18th Ave., Columbus, OH 43210, USA}
\affil{$^{6}$University of Canterbury, Department of Physics and Astronomy, Private Bag 4800, Christchurch 8020, New Zealand}
\affil{$^{7}$Korea University of Science and Technology, Daejeon 34113, Korea}
\affil{$^{8}$Department of Physics, Chungbuk National University, Cheongju 28644, Republic of Korea}
\affil{$^{9}$ Harvard-Smithsonian CfA, 60 Garden St., Cambridge, MA 02138, USA}
\affil{$^{10}$Jet Propulsion Laboratory, California Institute of Technology, 4800 Oak Grove Drive, Pasadena, CA 91109, USA}
\affil{$^{11}$Physics Department and Tsinghua Centre for Astrophysics, Tsinghua University, Beijing 100084, China}
\affil{$^{12}$Department of Physics, Zhejiang University, Hangzhou, 310058, China}
\affil{$^{13}$Canadian Institute for Theoretical Astrophysics,  University of Toronto, 60 St George Street, Toronto, ON M5S 3H8, Canada}
\affil{$^{14}$School of Space Research, Kyung Hee University, Yongin, Kyeonggi 17104, Korea}
\affil{$^{15}$Dipartimento di Fisica "E.R. Caianiello", Universit{\`a} di Salerno, Via Giovanni Paolo II 132, 84084, Fisciano, Italy} 
\affil{$^{16}$Istituto Nazionale di Fisica Nucleare, Sezione di Napoli, Napoli, Italy} 
\affil{$^{17}$Centre for Exoplanet Science, SUPA, School of Physics \& Astronomy, University of St Andrews, North Haugh, St Andrews KY16 9SS, UK} 
\affil{$^{18}$Astronomisches Rechen-Institut, Zentrum f{\"u}r Astronomie der Universit{\"a}t Heidelberg (ZAH), 69120 Heidelberg, Germany}
\affil{$^{19}$Niels Bohr Institute \& Centre for Star and Planet Formation, University of Copenhagen, {\O}ster Voldgade 5, 1350 Copenhagen, Denmark} 
\affil{$^{20}$Unidad de Astronom{\'{\i}}a, Universidad de Antofagasta, Av.\ Angamos 601, Antofagasta, Chile} 
\affil{$^{21}$Centre for Astrophysics \&\ Planetary Science, The University of Kent, Canterbury CT2 7NH, UK} 
\affil{$^{22}$Universit{\"a}t Hamburg, Faculty of Mathematics, Informatics and Natural Sciences, Department of Earth Sciences, Meteorological Institute, Bundesstra\ss{}e 55, 20146 Hamburg, Germany} 
\affil{$^{23}$Department of Astronomy, Stockholm University, Alba Nova University Center, 106 91, Stockholm, Sweden} 
\affil{$^{24}$Astrophysics Group, Keele University, Staffordshire, ST5 5BG, UK}
\affil{$^{25}$Institute for Advanced Research, Nagoya University, Furo-cho, Chikusa-ku, Nagoya, 464-8601, Japan}
\affil{$^{26}$Instituto de Astronomia y Ciencias Planetarias de Atacama, Universidad de Atacama, Copayapu 485, Copiapo, Chile} 
\affil{$^{27}$ Department of Physics, University of Rome Tor Vergata, Via della Ricerca Scientifica 1, I-00133—Roma, Italy} 
\affil{$^{28}$INAF -- Astrophysical Observatory of Turin, Via Osservatorio 20, I-10025 -- Pino Torinese, Italy}
\affil{$^{29}$CITEUC -- Center for Earth and Space Research of the University of Coimbra, Geophysical and Astronomical Observatory, R. Observatorio s/n, 3040-004 Coimbra, Portugal}
\affil{$^{30}$Department of Physics, Sharif University of Technology, PO Box 11155-9161 Tehran, Iran} 
\affil{$^{31}$Instituto de Astrof\'isica, Pontificia Universidad Cat\'olica de Chile, Av. Vicu\~na Mackenna 4860, 7820436 Macul, Santiago, Chile}
\affil{$^{32}$Centre for Electronic Imaging, Department of Physical Sciences, The Open University, Milton Keynes, MK7 6AA, UK} 
\affil{$^{33}$School of Physical Sciences, Faculty of Science, Technology, Engineering and Mathematics, The Open University, Walton Hall, Milton Keynes, MK7 6AA, UK} 
\affil{$^{34}$Stellar Astrophysics Centre, Department of Physics and Astronomy, Aarhus University, Ny Munkegade 120, 8000 Aarhus C, Denmark} 
\affil{$^{\dag}$NASA Postdoctoral Program Fellow}

\begin{abstract} 
We report the discovery of a cold Super-Earth planet 
($m_p=4.4\pm 0.5\,M_\oplus$) orbiting a low-mass ($M=0.23\pm 0.03\,M_\odot$)
M dwarf at projected separation $a_\perp = 1.18\,\pm 0.10$\,\au, i.e.,
about $1.9$ times the snow line.  The system is quite nearby
for a microlensing planet, $D_L=0.86\pm 0.09\,\kpc$.  Indeed, it was
the large lens-source relative parallax $\pi_\rel =1.0\,\mas$
(combined with the low mass $M$) that gave rise to the large,
and thus well-measured,
``microlens parallax'' $\pi_\e \propto (\pi_\rel/M)^{1/2}$ that
enabled these precise measurements.  OGLE-2017-BLG-1434Lb is the eighth
microlensing planet with planet-host mass ratio $q<1\times 10^{-4}$.
We apply a new planet-detection sensitivity method, which is
a variant of ``$V/V_\max$'', to seven of these eight planets
to derive the mass-ratio function in this regime.  We find
$dN/d\ln q\propto q^p$, with $p=1.05^{+0.78}_{-0.68}$, which confirms
the ``turnover'' in the mass function found by \citet{suzuki16}
relative to the power law of opposite sign $n=-0.93\pm 0.13$ at
higher mass ratios $q\ga 2\times 10^{-4}$.  We combine our result
with that of \citet{suzuki16} to obtain $p=0.73^{+0.42}_{-0.34}$.

\end{abstract}


\section{{Introduction}
\label{sec:intro}}

Microlensing searches are finding planets with 
planet/host mass ratio $q$ that are nearly uniformly distributed in
$\log q$ over $-4.3\la\log q \la -2$ (Figures 7 and 8 of \citealt{ob160596}).
Since planet detectability grows with $q$, this immediately implies
a steeply rising mass function toward lower mass ratios.  In the first
study to measure this relation, \citet{ob07368} found 
$d N/d\log q\propto q^n$ with $n=-0.7\pm 0.2$.  
Naively, the almost perfectly flat
distribution cataloged by \citet{ob160596} over a 2.3 dex range in $q$
would indeed appear to argue for a single power law.  However,
\citet{suzuki16} subsequently argued for a break in the power law
at about $\log q\sim -3.75$, with $n=-0.93\pm 0.13$ above the break.
Below the break they found a sign reversal in the power law, 
$d N/d\log q\propto q^p$ with 
$p=0.6^{+0.5}_{-0.4}$, implying a true ``turnover'' at the break.
The main argument for this break is that existing planet surveys
had significant sensitivity to planets below the break, but found 
very few.  In particular, the Microlensing Observations in Astrophysics
(MOA) survey, which was the primary data set that they analyzed,
had only two planets with $\log q<-4$.  Note, however, that another
recent study by \citet{shvartzvald16}, based on the overlap of the
OGLE, MOA and Wise surveys, fit to a single power law and 
found $n=-0.50\pm 0.17$.

A confirmation and refined measurement (or, alternatively, refutation)
of the \citet{suzuki16} 
break and turnover would be of great interest to constrain
theories of planetary formation.  Moreover, it would also be of
immediate practical interest in devising strategies for {\it WFIRST}
microlensing observations \citep{spergel13}.  Since {\it WFIRST} will
be far more sensitive to planets below the \citet{suzuki16} break
than ground-based surveys, its planet discovery rate is much more
sensitive to such a break.

Here, we report the discovery of the low mass ratio $q=5.8\times 10^{-5}$
planet OGLE-2017-BLG-1434Lb. This is the eighth microlensing planet
with a mass ratio that is securely in the range $\log q<-4$, 
meaning that the sample that lies clearly below
the \citet{suzuki16} break is now large enough for robust statistical
investigation.  On the other hand, in contrast to the \citet{suzuki16}
sample, these eight planets are drawn from quite heterogeneous detection
processes.  Therefore proper statistical analysis requires great care.

There are two other notable features of the discovery of OGLE-2017-BLG-1434Lb. 
First, we are able to make an accurate mass measurement, thanks to a
clear detection of the microlens parallax parameter,
\begin{equation}
\bpi_\e \equiv \pi_\e{\bmu_\rel\over\mu_\rel},
\qquad
\pi_\e \equiv {\pi_\rel\over\theta_\e};
\label{eqn:piedef}
\end{equation}  
where $\pi_\rel\equiv \au(D_L^{-1}-D_S^{-1})$ and $\bmu_\rel$
are the lens-source relative parallax and proper motion, respectively, 
\begin{equation}
\theta_\e\equiv \sqrt{\kappa M\pi_\rel};
\qquad
\kappa\equiv {4G\over c^2\au}\simeq 8.14{\mas\over M_\odot},
\label{eqn:thetaedef}
\end{equation}
and $M$ is the lens mass.  Note that when $\pi_\e$ and $\theta_\e$
are both measured, $M=\theta_\e/\kappa\pi_\e$ can likewise be determined
\citep{gould92,gould00,gould04,gouldhorne}.

Of the eight
planets with $\log q<-4$, five have good-to-excellent mass measurements,
which is a remarkably high fraction.  Two of these 
(including OGLE-2017-BLG-1434) have
excellent ground-based parallax measurements, one has a good ground-based
parallax measurement, one has an excellent 
space-based parallax measurement using {\it Spitzer},
and for the last, the host was directly imaged
8.2 years after the event.  This high rate of mass measurements  
permits at least a qualitative statement about the distribution of
host masses of low mass-ratio planets.

Second, although OGLE-2017-BLG-1434 is within a factor 1.3 of the
lowest mass-ratio microlensing planet, we show that it would have
been both detected and well-characterized even if it were a factor
30 smaller in $q$, i.e., $\log q = -5.71$, or approximately 12 Moon masses.
This demonstrates that
microlensing studies, at least in their current configuration, 
can probe to substantially smaller masses
than have yet been reported as discoveries, and hence further motivates
an investigation of whether there is really a break in the mass-ratio
function near $\log q\sim -4$.

\section{{Observations}
\label{sec:obs}}

OGLE-2017-BLG-1434 is at (RA,Dec)$_{J2000}$ = (17:53:07.29,$-30$:14:44.6),
corresponding to $(l,b)=(-0.28,-2.07)$.  
It was discovered and announced as a probable microlensing event
by the OGLE Early Warning
System \citep{ews1,ews2} at UT 19:33 25 Jul 2017\footnote{The same
microlensing event triggered an alert (OGLE-2017-BLG-1392) on 
a different catalog star $1.08^{\prime\prime}$ from OGLE-2017-BLG-1434
 at UT 19:14 20 Jul, i.e., 5.0 days earlier.  We find that the source
star lies at (RA,Dec) = (17:53:07.25,$-30$:14:44.5), which is roughly half way
between these two catalog stars, but slightly closer to OGLE-2017-BLG-1434.}.
The event lies
in OGLE field BLG501, for which OGLE observations
were at a cadence of $\Gamma = 1\,{\rm hr^{-1}}$ during the 2017 season
using their 1.3m telescope
at Las Campanas, Chile.

The Korea Microlensing Telescope Network (KMTNet, \citealt{kmtnet})
observed this field from its three 1.6m
telescopes at CTIO (Chile, KMTC), SAAO (South Africa, KMTS) and SSO 
(Australia, KMTA),
in its two slightly offset fields BLG01 and BLG41, with combined
cadence of $\Gamma = 4\,{\rm hr^{-1}}$ .

The great majority of these survey observations were carried out in $I$ band 
with occasional $V$ band observations made
solely to determine source colors.
All reductions for the light curve
analysis were conducted using variants of difference image analysis (DIA,
\citealt{alard98}), specifically \citet{wozniak2000} and \citet{albrow09}.

The MiNDSTEp collaboration observed OGLE-2017-BLG-1434 from the 1.54m Danish
Telescope at La Silla, Chile, 
using an
EMCCD camera operated at 10 Hz and with a broad passband approximating 
the $I$-band filter, with
mean response at 770 nm \citep{skottfelt15,evans16}. 
The data were reduced using an updated version of the
DanDIA pipeline \citep{bramich08}.
These observations were initiated with
10 exposures on the night of 
${\rm HJD}^\prime = 7981$ and continually increased to 75 exposures on
${\rm HJD}^\prime = 7984$, i.e., the night closest to the predicted peak,
falling to 50 exposures the following night.  Based on these last observations,
MiNDSTEp issued an alert, triggered by SIGNALMEN \citep{signalmen},
 of a possible anomaly, which it confirmed
the next day based on OGLE online data.  MiNDSTEp then continued regular
observations of the event until the night of ${\rm HJD}^\prime = 7998$.



\section{{Analysis}
\label{sec:analysis}}

\begin{figure}
\vspace{-3.5cm}
\plotone{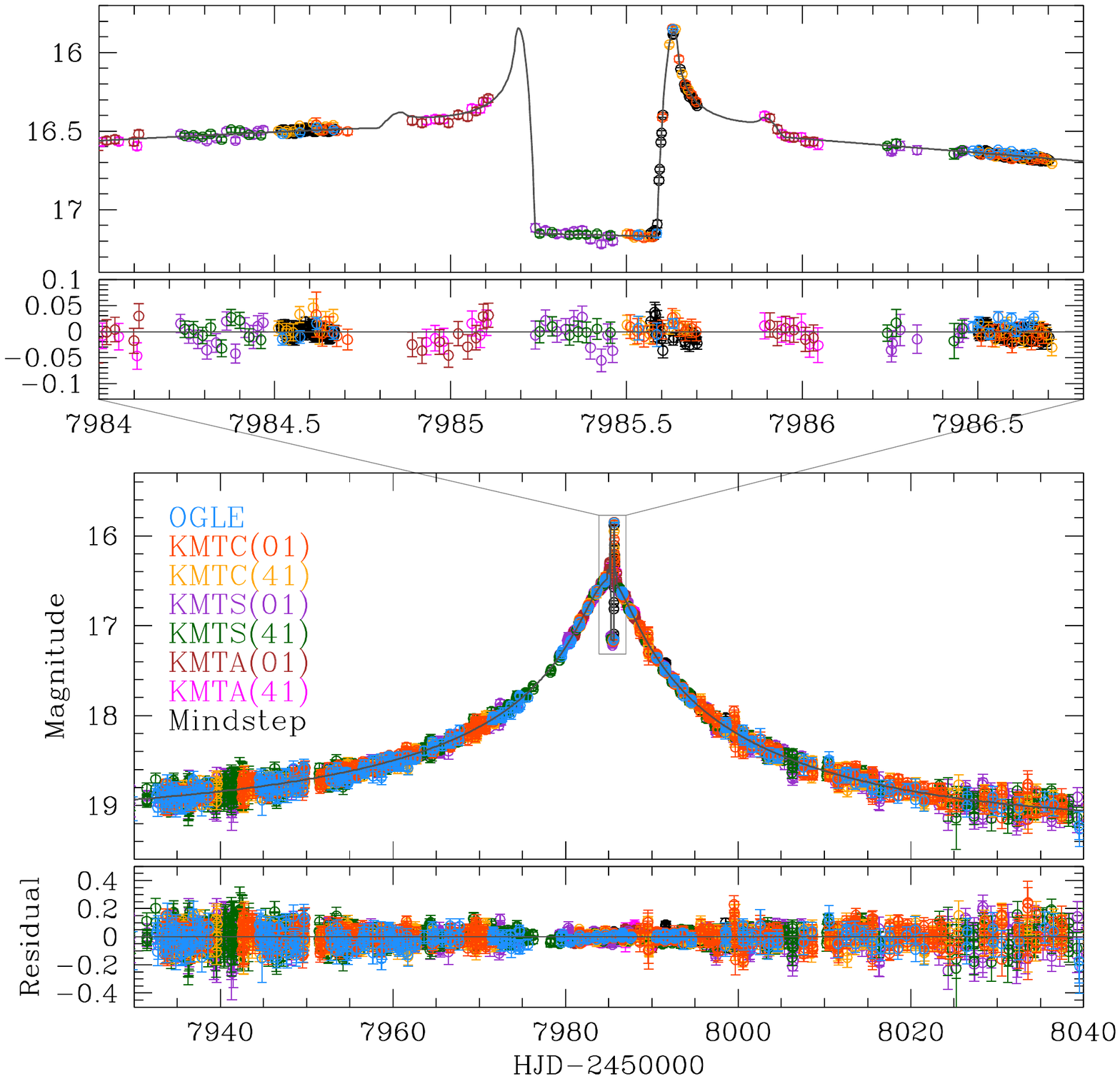}
\vspace{-3cm}
\caption{Light curve and best-fit model of OGLE-2017-BLG-1434.
As discussed in Section~\ref{sec:analysis}, many of the key parameters
can be ``read off'' the light curve, including that this is a very
low mass-ratio planet: $q<10^{-4}$.  Data are color-coded by observatory.
}
\label{fig:lc}
\end{figure}

With the exception of some deviations near the peak that last slightly more
than one day, the overall shape of the OGLE-2017-BLG-1434 light curve
(Figure~\ref{fig:lc}) is that of an ordinary point-lens \citep{pac86} event.
Excluding the deviation, the change in magnitude from baseline
to peak ($\sim 2.5$ mag) implies a magnification $A \ga 10$ (higher
if there is significant blending).
The two most obvious components of the deviation are a flat trough
that lies about 0.7 mag below the level of the point-lens curve and
lasts 0.4 days (traced in KMTS, OGLE, KMTC, and MiNDSTEp data),
followed by a very rapid caustic entrance (traced in KMTC, OGLE, and MiNDSTEp
data).  Once these features are noted, it becomes clear that the underlying
light curve is essentially symmetric and that the onset of
a caustic exit is traced by KMTA data just before the trough.

This trough is an unambiguous signature of a ``minor image perturbation''.
The point-lens light curve (due to the host in the absence of a companion)
is generated by two images of the source.  The major image forms at a
minimum in the time-delay surface (in accord with Fermat's principle),
while the minor image forms at a saddle point.  The latter is highly
unstable, so that if a small planet lies at or near the position of
this image, the image will be virtually annihilated.  The flux
ratio of the two images is $\eta = (A-1)/(A+1)$ where $A$ is the total
magnification.  Hence, for a high-magnification event such as this one,
$\eta\rightarrow 1$, which means that annihilating the minor image should
decrease the total flux by half, i.e. $\sim 0.75$ mag.  That is,
the light curve's behavior exactly corresponds to this expectation.

Such troughs are always aligned with the planet-star axis and are
flanked by two caustics.  However, the
troughs generally extend substantially beyond the caustics.  Hence,
if the source trajectory crosses the trough at a point where there are
no caustics, its entrance and exit to the trough will be smooth.
However, if it crosses the caustics, the trough entrance and exit
will correspond to a sharp (discontinuous slope)
caustic exit and entrance, respectively.
The latter is clearly the geometry of OGLE-2017-BLG-1434.

Of course, after exiting the trough (so, entering the caustic),
the source must again exit the caustic.  Because the caustic edge
facing the trough is much stronger than its ``outer walls'', the
effect on the light curve is much less pronounced. Nevertheless,
this outer-edge crossing was captured in the first three KMTA points 
after the trough.

\begin{figure}
\vspace{-4cm}
\plotone{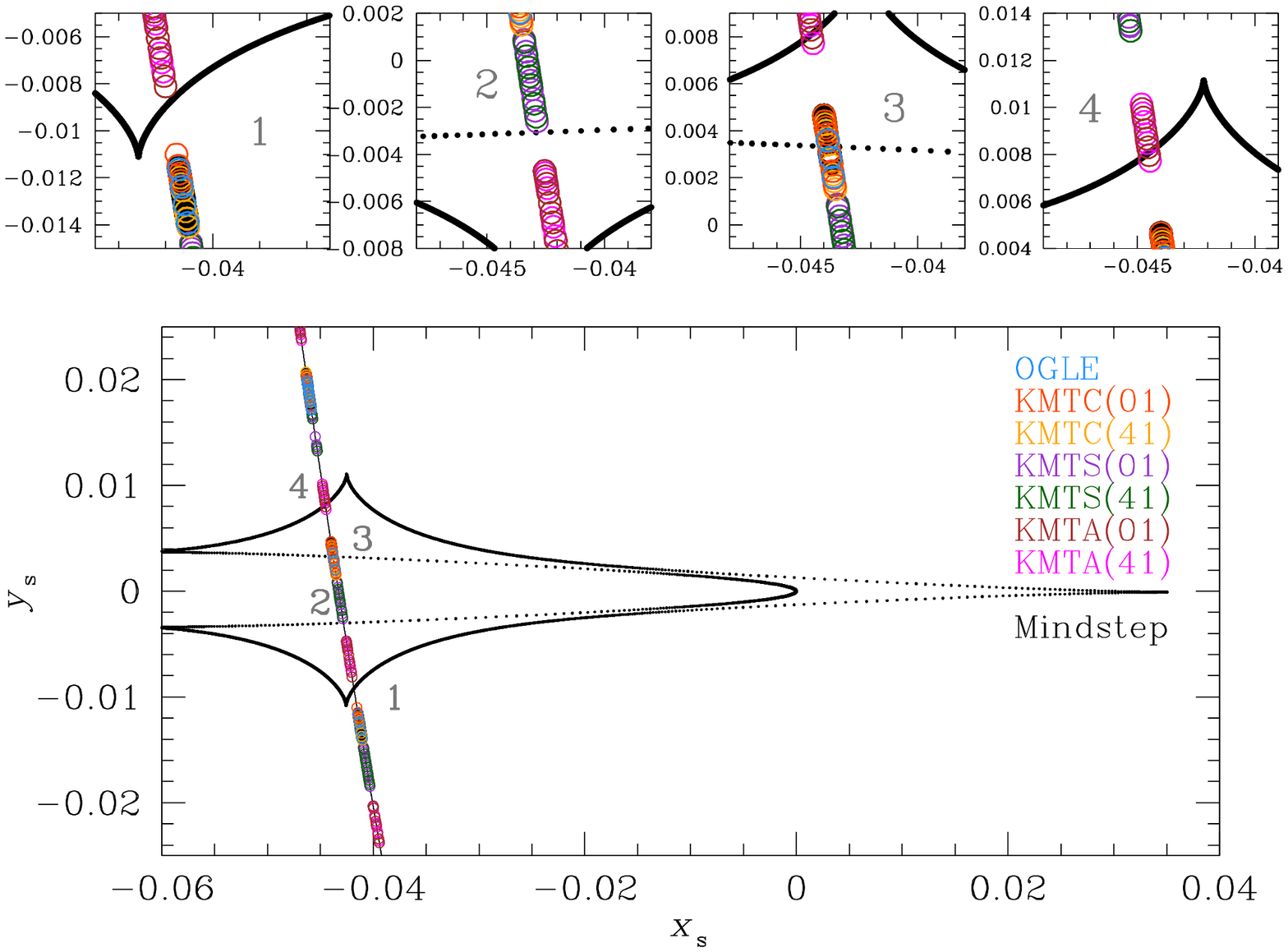}
\vspace{-3cm}
\caption{Caustic diagram of OGLE-2017-BLG-1434.  The source passes
over the ``planetary wing'' of a resonant caustic, resulting from the
planet perturbing the minor image.  The points are color coded by
observatory, and their size represents the scaled source
$\rho=4.5\times 10^{-4}$ of the best fit model.
}
\label{fig:caust}
\end{figure}

If the trough occurs at relatively low magnification, then each
of the two caustic ``walls'' that flank the caustic will be part
of triangular caustic structures.  However, at progressively
higher magnification (corresponding to planets that are progressively
closer to the Einstein ring), these triangular caustics grow in size
and progressively move toward the quadrilateral central caustic
close to the host.  The triangular caustics eventually merge with 
the central caustic to form a single,
six-sided caustic (Figure 4 of \citealt{gaudi12}).
This turns out to be the geometry of OGLE-2017-BLG-1434.
See Figure~\ref{fig:caust}. 


Continuing this logic, one can approximately read off
the parameters of a standard seven-parameter model from the light curve, 
using known analytic formula \citep{han06} for the caustics.  
Three of these parameters
$(t_0,u_0,t_\e)$ correspond to the time of maximum, impact parameter,
and Einstein crossing time of the underlying point-lens \citep{pac86}
model.  Three others $(s,q,\alpha)$ describe the binary companion,
namely its separation (in units of $\theta_\e$), its mass ratio, and
the angle of the binary axis relative to the source trajectory.
Finally, if (as in the present case) the source passes over or near
the caustics, one must specify $\rho=\theta_*/\theta_\e$, i.e., the
ratio of the source radius to the Einstein radius.  

A point-lens fit to the light curve with the anomaly removed yields
\hfil\break\noindent
$(t_0,u_0,t_\e)=(7984.94,0.027,95\,{\rm day})$,
which implies an effective timescale $t_\eff \equiv u_0 t_\e=2.57\,$ days.  
The anomaly is centered $\delta t= 0.5\,$days after peak, implying
that $\alpha={\rm atan}(-t_\eff/\delta t)= \pm 259^\circ$ and that
$s$ should satisfy, $s^{-1}-s= u_0\sqrt{1+(\delta t/t_\eff)^2}$, which
implies\footnote{The other $(s>1)$ solution is excluded because it
would correspond to the major image.}
$s\simeq 1-u_0/2 = 0.986$.  From the light curve,
the duration of the trough is $\Delta t\simeq 0.4\,$days.  (It could
in principle be slightly larger because the caustic exit is not actually
observed.  However, the rise toward the caustic is observed from KMTA,
so this estimate cannot be far off.)  This quantity can be related
to the \citet{han06} parameter $\eta_{c-} = 2q^{1/2}(s^{-2}-1)^{1/2}$
by $\Delta t = 2 t_\e\eta_{c-}|\sec\alpha|$, which for the present case implies
$q\simeq u_0(\Delta t/t_\eff)^2/16 = 4.1\times 10^{-5}$.
Finally, the rise time of the caustic exit
in OGLE/KMTC/MiNDSTEp data is $t_{\rm rise} = 66\,$min.  From
\citet{gouldandronov99}, $t_{\rm rise} = 1.7\,t_*\sec\alpha$, where
$t_* = \rho t_\e$.  Hence, $t_*=39\,$min and
$\rho \simeq (t_*/t_\e) =3.5\times 10^{-4}$. 

However, following from these results, it is obvious that additional
higher order effects should be measurable.
First note that $\rho$ is unusually small, so that the Einstein radius
$\theta_\e = \theta_*/\rho \simeq 3000\,\theta_*$.
We will estimate the source size $\theta_*$ in detail in Section~\ref{sec:cmd}.
However, just from the source flux derived from the \citet{pac86} fit, it
is an upper main sequence star, i.e., $\theta_*\sim 0.5\,\muas$.

Such a large Einstein radius $(\theta_\e\sim 1.5\,\mas)$
immediately implies that the host must
be either a dark remnant (black hole or neutron star) or it must
be quite nearby.  That is, from the definition of $\theta_\e$
(Equation~(\ref{eqn:thetaedef})),
\begin{equation}
\pi_\rel = {\theta_\e^2\over\kappa M} \rightarrow 0.3\,\mas
\bigg({M\over M_\odot}\biggr)^{-1} .
\label{eqn:thetaeeval}
\end{equation}
Indeed, since a solar-mass star at $D_L\sim 2.5\,\kpc$ would easily be
visible, the actual lens must have even lower mass (hence higher $\pi_\rel$).
Therefore, again unless the host is a dark remnant, the microlens
parallax must be fairly large.
\begin{equation}
\pi_\e = {\pi_\rel\over\theta_\e}> 0.2 .
\label{eqn:pieeval}
\end{equation}
Given that the event is quite long, such a parallax should be 
measurable.

Thus, without any detailed modeling, one can infer that there should
be a strong microlens parallax signal and that the implications of not
finding such a signal would be striking.

When introducing the two parallax parameters $\bpi_\e=(\pi_{\e,N},\pi_{\e,E})$,
one must also, at least initially, introduce linearized orbital motion
parameters $(ds/dt,d\alpha/dt)$ as well.  These encode the instantaneous
rate of change in the separation and orientation of the binary at $t_0$.
There are two reasons that these must be included.  First, the orbital
motion parameters $(ds/dt,d\alpha/dt)$ can be correlated with $\bpi_\e$,
so that by ignoring them one can induce artificial effects in the parallax
\citep{ob09020,mb09387,ob150768}.  Second, binary systems are known a priori
to orbit their center of mass.  Hence, there is no viable reason for
excluding these parameters except if they are better constrained
by the fact that physical systems ought to be bound than they are by the
data.  However, while (as described above) there are very strong
reasons to believe that the parallax parameters can be measured, there
is no corresponding confidence with respect to $(ds/dt,d\alpha/dt)$.
Therefore, these parameters must be handled carefully.  See, e.g.,
\citet{ob160693}.

Notwithstanding the above analytic arguments, we conduct a grid search
over $(s,q,\alpha)$, seeded by the above values of $(t_0,u_0,t_\e,\rho)$,
with all parameters except $(s,q)$ allowed to vary and apply $\chi^2$
minimization using a Markov Chain Monte Carlo (MCMC).  
To evaluate the magnifications at individual 
data points, we use inverse ray shooting in and near the caustics
\citep{kayser86,schneider88,wambs97} and 
multipole approximations \citep{pejcha09,gould08} elsewhere.
We employ a linear limb-darkening coefficient $\Gamma_I = 0.429$
based on the source type derived in Section~\ref{sec:cmd}.

We find only one solution. This is close to the one derived above 
analytically in terms of $(s,\alpha)$ and the so-called invariant quantities
\citep{mb11293,ob160693}:
$(s,\alpha,t_\eff,t_*,q t_\e)=[0.981,259^\circ,(2.59,0.0284,0.00364)\,{\rm day}]$
compared to\hfil\break\noindent
 $[0.986,259^\circ,(2.57,0.0271,0.00389)\,{\rm day}]$
``predicted''.  The major difference is only in $t_\e$,
which can be significantly impacted by unmodeled parallax for long
timescale events.

We then introduce the higher-order parameters $\bpi_\e=(\pi_{\e,N},\pi_{\e,E})$
and $\bgamma = ((ds/dt)/s,d\alpha/dt)$.  We find that in a completely
free fit, three of these are well constrained, but the fourth
$(\gamma_\perp=d\alpha/dt)$ is not.  In particular, we find that for
most of the solution space, the (absolute value of the)
ratio of projected kinetic to potential
energy,
\begin{equation}
\beta\equiv \biggl({\rm KE\over PE}\biggr)_\perp
= {\kappa\,M_\odot\,{\rm yr}^2\over 8\pi^2}\,{\pi_\e\over\theta_\e}\,
{s^3\gamma^2\over(\pi_\e + \pi_s/\theta_\e)^3},
\label{eqn:betadef}
\end{equation}
violates the boundedness condition, $\beta<1$.  Here, we adopt
$\pi_s=0.117\,\mas$ for the source parallax.  We address this by
making two different calculations.  First, we arbitrarily set $\bgamma=0$.
Of course, as mentioned above, this is unphysical, but it is simple and
is useful as a benchmark for the second calculation, in which we
allow $\bgamma$ to vary but restrict $\beta<0.7$, i.e., a limit that 
would be satisfied by the overwhelming majority of real, bound 
systems\footnote{The results hardly differ if we choose the extreme
physical limit $\beta<1$.}.

The results are shown in Table~\ref{tab:ulens}.  The first point is that the 
parallax+orbital models
in which $\beta$ is restricted to a reasonable physical range
$\beta<0.7$ yield statistically indistinguishable results from the
parallax-only models in which $\beta=\gamma=0$.  That is, our inability
to fully measure $\gamma$ does not significantly influence the
measurement of any other parameter.

Second, while there are two degenerate $(\pm u_0)$ models with
similar $\chi^2$, their parameters (apart from the sign
of $u_0$) are the same within $1\,\sigma$.  Hence, this degeneracy does
not materially impact the inferred physical parameters of the
system.

\section{Physical Parameters}
\label{sec:phys}


\subsection{Measurement of $\theta_\e$ and $\mu$}
\label{sec:cmd}

We derive the source brightness from the model presented in 
Section~\ref{sec:analysis} and derive the color from regression. We then find
the offset from the clump on an instrumental color-magnitude diagram:
$[(V-I),I]_s - [(V-I),I]_{\rm clump} = (-0.328,+3.990)\pm (0.023,0.038)$.
We adopt $[(V-I),I]_{\rm clump} = (1.06,14.46)$ from \citet{bensby13} and
\citet{nataf13}, respectively, and so derive $[(V-I),I]_s =
(0.732,18.45)\pm (0.025,0.063)$.  We convert from $V/I$ to $V/K$
using the $VIK$ color-color relations of \citet{bb88} and finally
derive
\begin{equation}
\theta_* = 0.657\pm 0.041\,\muas
\label{eqn:thetastar}
\end{equation}
using the color/surface-brightness relations of \citet{kervella04}.
Incorporating parameters from Table~\ref{tab:ulens}, we thereby derive.
\begin{equation}
\theta_\e ={\theta_*\over\rho} = 1.40\pm 0.09\,\mas;
\qquad
\mu ={\theta_*\over t_\e} = 8.1\pm 0.5\,\masyr .
\label{eqn:thetaemu}
\end{equation}

\subsection{Masses, Distance, and Projected Separation}
\label{sec:mdps}

 Combining the results of Section~\ref{sec:cmd} and Table~\ref{tab:ulens}, 
we find
\begin{equation}
M = {\theta_\e\over\kappa\pi_\e} = 0.234\pm 0.026\,M_\odot;
\qquad
m_p = {q\over 1+q}M = 4.4\pm 0.5\,M_\oplus
\label{eqn:mhostmplanet}
\end{equation}
and
\begin{equation}
D_L = {\au\over\theta_\e\pi_\e + \pi_s}  = 0.86\pm 0.09\,\kpc
\qquad
a_\perp = s\theta_\e D_L = 1.18\pm 0.10\,\au ,
\label{eqn:dlaperp}
\end{equation}
where we have adopted a source parallax $\pi_s = 0.117\pm 0.010\,\mas$
and where $a_\perp$ is the projected separation.  That is, the
planet is a super-Earth orbiting a middle-late M dwarf.  If
we adopt a ``snow line'' scaled to host mass (e.g., \citealt{kennedy08}),
and anchored in the observed Solar-system value,
$a_{\rm snow} = 2.7\,\au\,(M/M_\odot)$, then this planet lies
projected at $a_\perp = 1.9\,a_{\rm snow}$.

\section{Microlensing Earths and Super-Earths with Well-Measured Masses}
\label{sec:massmeas}

OGLE-2017-BLG-1434Lb joins a small list of Earths and
Super-Earths with well-measured masses discovered by microlensing.
To be an ``Earth or Super-Earth'', we require a best-estimated
planet mass $m_p<7\,M_\oplus$.
To be ``well-measured'' we set two requirements.  First
we require that the quoted $1\,\sigma$ error on the planet mass measurement
span a factor $<2$, i.e., $\sigma(\log(m_p))<0.15$.  Second, we
require that the host mass was determined either by measuring both $\pi_\e$
and $\theta_\e$ (as was done here) or by directly imaging the host.

We review the literature 
(effectively updating the summary by \citealt{ob160596})
and find only three such planets:
OGLE-2016-BLG-1195Lb \citep{ob161195a,ob161195b},
OGLE-2013-BLG-0341LBb \citep{ob130341},
and
OGLE-2017-BLG-1434Lb (this work).  In all cases, the mass
determination is via measurements of $\theta_\e$ and $\pi_\e$.
We note that there is another planet, MOA-2007-BLG-192Lb 
\citep{mb07192,mb07192k} whose host mass was quite well determined by direct
imaging and whose best-estimated planet mass $m_p=3.2_{-1.8}^{+5.2}$
falls in the defined range.  However, the error bars on the planet
mass are far too large to meet our criterion.

Table~\ref{tab:lowm} gives the main characteristics of these systems.


The first point to note about these three well-constrained low-mass
planets is that they all have
low-mass hosts, i.e., from the hydrogen-burning limit to a middle-late
M dwarf.  The second point is that all are seen projected close to the 
Einstein ring, $s\sim 1$.  And the third is that two of the three are 
extremely nearby, $D_L\la 1\,\kpc$.  
All three of these characteristics are heavily influenced 
by selection, but none can be regarded purely as a selection effect.

\begin{figure}
\vspace{-6cm}
\plotone{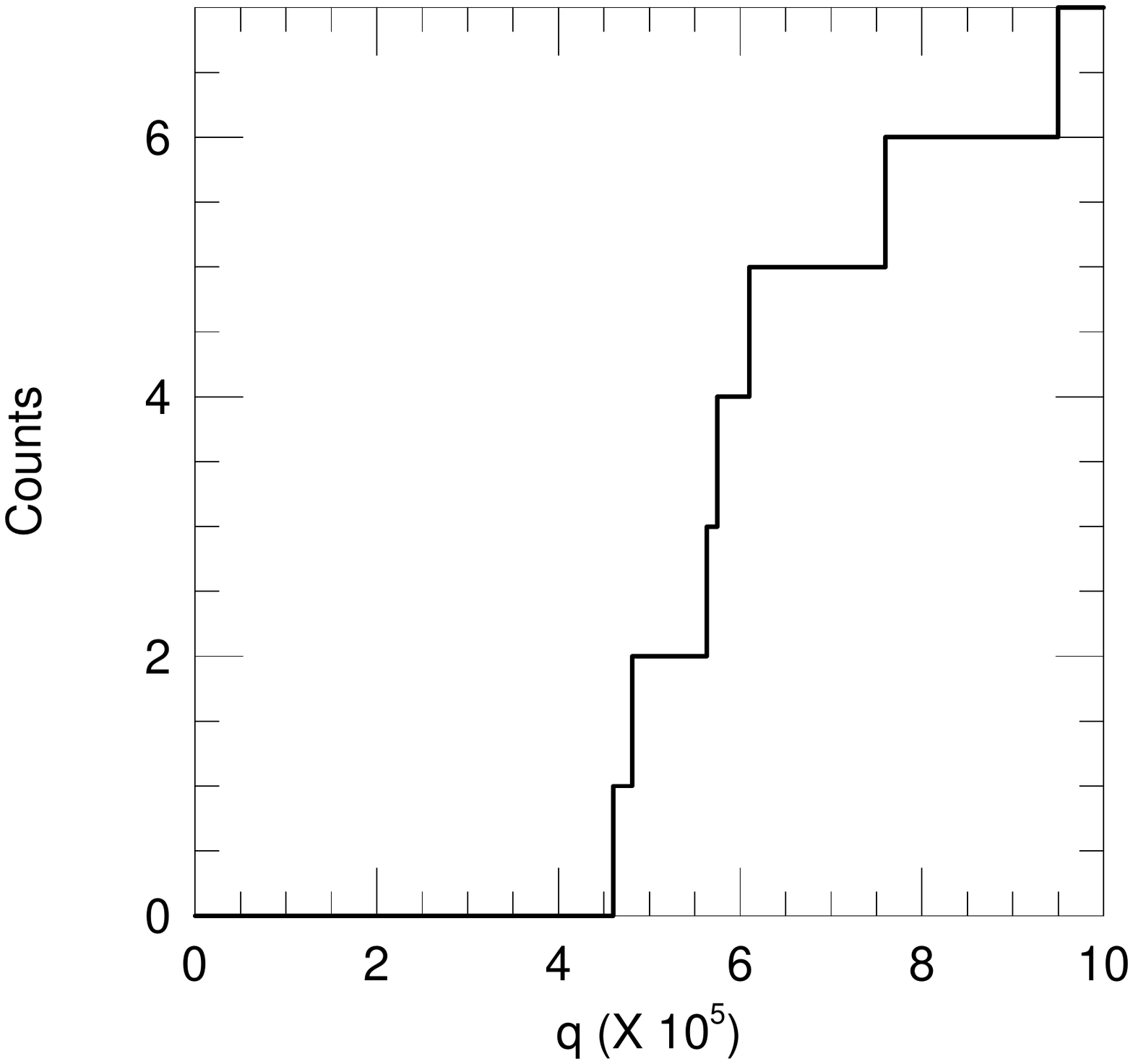}
\caption{Cumulative distribution of planet/host mass ratios $q$ for
the seven microlens planets with well-defined measurements $q<10^{-4}$.
Five of the seven have $4.6\times 10^{-5}\leq 6.1\times 10^{-5}$,\
suggesting either a rapid drop either in sensitivity of microlensing
experiments to low mass-ratio planets or in the frequency of such planets.
}
\label{fig:qlist}
\end{figure}

As illustrated by 
Figure~\ref{fig:qlist}, there are no microlensing planets with 
$q\la 5\times 10^{-5}$.  This fact, combined with our sample 
requirement $m_p<7\,M_\oplus$, already implies that the hosts 
will be $M\la 0.4\,M_\odot$.  However, as we will show in
Section~\ref{sec:massrat}, the apparent ``barrier'' at $q\sim 5\times 10^{-5}$
is not a selection effect: lower mass ratios could have been detected
if such planets were common.

Similarly, it is well-known that it is easier to detect planets when
they lie projected close to the Einstein ring.
In particular, for relatively high-magnification events, which applies
to all three of these planets, planet-sensitivity diagrams have a triangular
shape that is symmetric about $\log s = 0$ \citep{gould10}.  However,
since (as just mentioned) planets could have been detected at lower $q$,
it follows immediately that they could also have been detected at
higher $|\log s|$.

Finally, ground-based parallax measurements are heavily biased toward
nearby lenses, simply because the microlens parallax is larger:
$\pi_\e=(\pi_\rel/\kappa M)^{1/2}$.  However, while this bias is quite
strong for bulge lenses (to the point that there are no ground-based
parallax measurements for events with unambiguous bulge lenses), it is only
moderately strong for events of intermediate (few kpc) distance, 
and there are many
ground-based parallax measurements for events at intermediate distances.

Thus, although this sample is heavily affected by selection, it does
contain some information.  Nevertheless, given that this sample is
both small and biased by selection, we refrain from using it to
draw any systematic conclusions.

\section{Planet Mass-Ratio Function at the Low-Mass End}
\label{sec:massrat}

At $q=5.8\times 10^{-5}$, OGLE-2017-BLG-1434Lb is the eighth published 
microlensing planet with a planet/host mass-ratio measurement that places
it securely in the range $q<10^{-4}$.  Strikingly, these are mostly 
clustered close to 
$q\sim 5\times 10^{-5}$. 
See Figure~\ref{fig:qlist}.
This would seem to indicate a sharp cutoff either in sensitivity or
in the existence of planetary companions at the separation ranges
accessible to microlensing.  

Because the discovery process of these
planets is quite heterogeneous, it is not possible to reliably
determine an absolute mass-ratio function from this sample.
That is, there is no way to estimate the rate of non-detections
from which this sample was drawn.

However, by applying the technique of ``$V/V_\max$'' \citep{schmidt68},
we can use
this sample to constrain the relative frequency by mass ratio.
That is, we can consider various trial mass-ratio functions $F(q)$.  For
each detected planet $i$, we evaluate the ``$V/V_\max$'' ratio $r_i$
defined by
\begin{equation}
r_i = {\int_{q_i}^{q_\max} d\ln q^\prime F(q^\prime) P_i(q^\prime)
\over\int_{0}^{q_\max} d\ln q^\prime F(q^\prime) P_i(q^\prime)},
\label{eqn:vvmax}
\end{equation}
where $q_\max=10^{-4}$ (i.e., the definition of the sample) and
$P_i(q^\prime)$ is the probability that the planet would have been
detected and published if the event had had exactly the same
parameters as the actual one, but with a different $q^\prime\not= q_i$.

If $F(q)$ has been chosen correctly, then the distribution
of the $r_i$ should be consistent with being drawn from a uniform
distribution over the interval [0,1].  Hence, all trial mass-ratio
functions $F(q)$ that yield a distribution of $r_i$ that is inconsistent
with uniform can be ruled out.

In principle, one might consider each $P_i$ to be a continuous function
varying between zero and one.  For example, one might decide that
$P_3(q^\prime=1.3\times 10^{-5})=43\%$.  That is, the light curve associated
with the third event (and with the specified $q^\prime$) would have had a
43\% chance of having been noticed as planetary in nature and then
generating sufficient confidence in the evaluation to publish it.
In fact, we will approximate the $P_i(q)$ as bi-modal, either 0 or 1.
In most cases, this means that there is some continuous interval
over which the planet is judged to be detectable, defined by $q_{\min,i}$.
Then Equation~(\ref{eqn:vvmax}) would become
\begin{equation}
r_i \rightarrow {\int_{q_i}^{q_\max} d\ln q^\prime F(q^\prime) 
\over\int_{q_{\min,i}}^{q_\max} d\ln q^\prime F(q^\prime)}.
\label{eqn:vvmaxspec}
\end{equation}
However, as we will show, there is one event for which the
range of discoverable planets could in principle have been
discontinuous, so we
retain the more general form of Equation~(\ref{eqn:vvmax}), at least
at the outset.

\subsection{Evaluation of $q_{\min,i}$ and $P_i(q^\prime)$}
\label{sec:Pqeval}

We define our sample by the following three criteria:
\begin{enumerate}
\item[](1) Best-fit mass ratio $\log q< -4$,
\item[](2) Formal error estimate $\sigma(\log q)<0.15$,
\item[](3) No alternate solutions with $\Delta\chi^2<10$ and $\Delta\log q>0.3$ .
\end{enumerate}
Criterion (1) is the regime that we seek to probe. Planetary candidates
that fail criterion (2) generally cannot be securely identified as being in
the sample.  Moreover, for planets with larger error bars, there is
an increasing (and basically unknowable) probability that they would
not be published.
Candidates that fail criterion (3) are ambiguous in the
sense that $q_{\min}$ can be substantially different for different solutions.

For consistency with these choices, we likewise set $P_i(q^\prime)=0$
for any choice of $q^\prime$ that leads 
to failure of either criterion (2) or (3).

We find that one of the eight planets that satisfy criteria (1) and (2),
fails criterion (3): OGLE-2017-BLG-0173 \citep{ob170173}.  In fact,
at the time that we devised these criteria, it was not yet known that
OGLE-2017-BLG-0173 suffered from such a degeneracy.  More generally,
we did not alter our criteria as we studied the eight planets in 
detail.  Even though one of these, OGLE-2017-BLG-0173, will be excluded 
from the sample, we include it in this part of the analysis for 
completeness and to explore the problems posed by this type of degeneracy.

Here we evaluate the range of $q$ over which each of the eight planets
that satisfy criteria (1) and (2) would have been detected.  In
each case we discuss the methods by which the planet was detected, or
could have been detected using approaches that were applied to
essentially all events in its class.  ``Detection'' here requires 
that a simulated planet must meet two criteria:  first,
that it would have been noticed as a potential planet based on
whatever data were routinely available, and second, that further
analysis based on re-reduced data (plus whatever archival data would
have been available) would have led to an unambiguous planet, worthy
of publication.  Because all the planets discussed here have low mass
ratio $q$, we assume that if the planet was publishable, it would in fact
have been published.  (This is not actually true for some higher
mass-ratio planets, which sometimes take many years to elicit enough
enthusiasm to push through to publication.)

\subsubsection{OGLE-2017-BLG-1434}
\label{sec:ob171434}

OGLE-2017-BLG-1434 was first noticed as a potential planetary event
by the MiNDSTEp collaboration based on its own data, as described in
Section~\ref{sec:obs}.  If the mass ratio $q$ had been somewhat
lower, then MiNDSTEp data would have still shown a strong anomaly, and
a similar alert would have almost certainly been issued.  However,
our concern here is to identify the lowest mass ratio $q$ that
would have yielded a noticeable signature, which would then trigger
further investigation.  As we show below, at sufficiently low
$q$, there would have been no noticeable deviations as seen from Chile,
but there would still have remained 
significant deviations as seen from South Africa.
In these limiting cases, there would have been no MiNDSTEp alert.
The path to anomaly recognition would then have been the regular
KMTNet review of ongoing events.  This review would have combined
OGLE online data with KMTNet ``quick look'' data.  Hence, this is what
we simulate below.

\begin{figure}
\vspace{-3.5cm}
\plotone{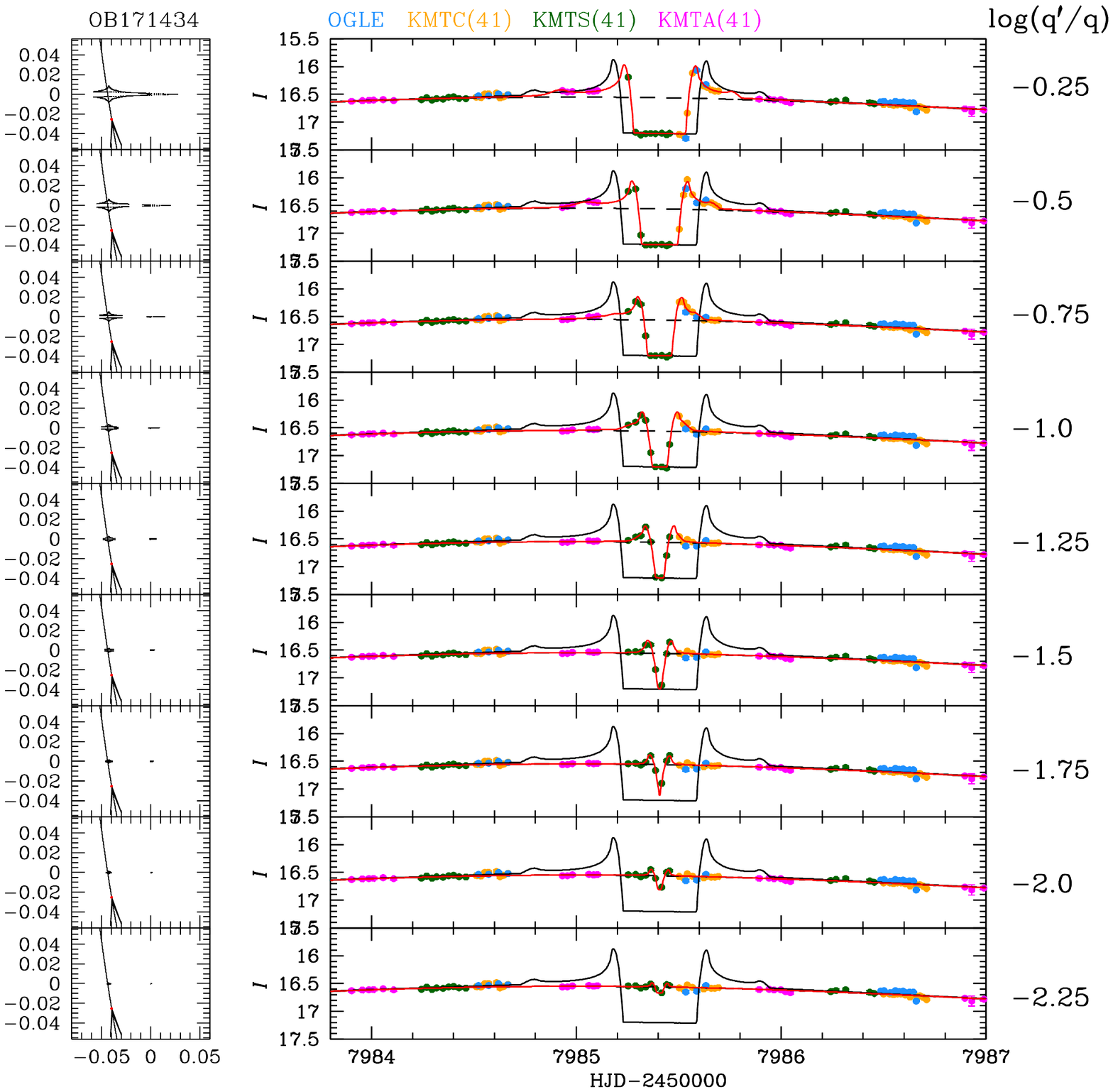}
\vspace{-3cm}
\caption{Nine simulations of OGLE-2017-BLG-1434 with exactly the same
parameters as the best-fit model (black curve) except that the mass ratio $q$ is
lower by $\Delta\log q$ as indicated in the right axis labels.  In each
case, the simulated data points (various colors) deviate from the model 
(orange curve) by exactly the same amount as the actual data points deviate
from the best-fit model.  The left panels show the corresponding caustic
geometries.  These characteristics will be same for
all eight events in the figures that follow.  In this figure, the
data points are based on the ``online'' OGLE data and ``quick look''
KMTNet data in order to focus on the problem of determining whether
the event would be recognized as sufficiently interesting to trigger
re-reductions.
}
\label{fig:ob171434}
\end{figure}

Figure \ref{fig:ob171434} 
shows the anomalous region of the light curve in on-line
OGLE data and quick-look KMTNet data as it would have appeared
with exactly the same parameters shown in Table~\ref{tab:ulens}, except with 
$q$ taking on other values.  To construct this figure, we measure
the residuals from the best-fit model for each data point and
renormalize the errors (with uniform error rescaling) so that
$\chi^2/dof=1$.  Then for each observatory, we create a fake light curve
whose value is the model light curve plus the residual in magnitudes,
and whose error bar is the same as that of the original (renormalized) 
data point.  In each case, we show both the original model and the
model with different $q$ that was used to construct the fake light
curves.  Note that the KMTNet ``quick look'' data consisted only
of observations from field BLG41 (and not BLG01) and so are 
at half the cadence of the final KMTNet data that are 
shown in Figure~\ref{fig:lc}.

The figure has nine panels corresponding to 
$\log(q^\prime/q)=-0.25,-0.5,\ldots -2.25$.  Careful inspection of the
$\log(q^\prime/q)=-1.75$ panel shows clear systematic residuals,
with two consecutive points below the point-lens curve, with the
lower one being below by $\sim 0.3\,$mag.  However,  a cursory inspection
would have shown only a single clearly outlying point.
Based on the direct experience 
of the KMTNet team, we consider that it is possible that 
these would have triggered a further investigation, i.e., first
corroboration in the BLG01 data and, following this,
re-reduction of all of the data.  However, we judge that the probability
for this is substantially below 50\%, and hence within the framework
we have adopted, we approximate this probability as $P=0$. By contrast,
the ``quick look'' data for $\log(q^\prime/q)=-1.50$, with two points
well below the single-lens curve, and one of them by $\sim 0.6\,$mag
would certainly have triggered such an investigation.
On the other hand, $\log(q^\prime/q)=-2.0$ would certainly
not have triggered such an investigation, but even it had, the
re-reduced data would not have yielded a publishable detection because
the signal is too weak.

To assess publishability, we first consider the marginally 
recognizable case, $\log(q^\prime/q)=-1.75$.
We model the fake re-reduced data in exactly the same way that
we modeled the real data except that we exclude orbital motion
(which would not be measurable for such a short and weak anomaly,
and also would not be at all required for publication).  We find that
the planet's parameters in this case are well constrained, for example
$\log q=-5.88\pm 0.072$.  This error bar is well within the limit
set by criterion (2).  We note that the best fit value ($-5.88$) differs
by $1.4\,\sigma$ from the ``input value'' of the simulated data ($-5.98$).
This difference reflects the fact that the residuals, which concretely
reflect the observational errors, are preserved in the simulated data.
It follows that the much stronger signal at $\log(q^\prime/q)=-1.50$
would also meet our criteria.  We therefore adopt this threshold.

Before continuing, we note that more systematic procedures are
currently being applied to 2017 KMTNet data, by means of
which it is very likely that at $\log(q^\prime/q)=-1.75$, this planet
would ultimately have been discovered.  That is, while the quick-look
data were restricted to BLG41, all known microlensing events
(whether discovered by KMTNet or others) will by mid-2018 be
reviewed using all available KMTNet data.
However, such ``ultimate discoveries'' are irrelevant to
the analysis being conducted here.  
All real planets that will ``ultimately
be discovered'' are presently unknown, and thus are automatically
excluded.  Therefore, we must equally exclude simulated planetary
events in this class. 


\subsubsection{OGLE-2017-BLG-0173}
\label{sec:ob170173}
\citet{ob170173} analyzed OGLE and KMTNet data for OGLE-2017-BLG-0173 and found
three solutions, including two ``von Schlieffen'' solutions (A,C) with
$q\sim 6.5\pm 0.9\times 10^{-5}$, and one ``Cannae'' solution (B) with
$q\sim 2.5\pm 0.2\times 10^{-5}$.  All three solutions have 
$s\sim 1.5$.  Two of these solutions (A,B) differ by only $\Delta\chi^2=3.5$,
so this planet fails criterion (3), even though it satisfies 
criteria (1) and (2).  Hence, this planet is excluded from our sample.
Nevertheless, as mentioned above, we analyze it for completeness.
Since solution (C) has $\Delta\chi^2=16$, we focus here on solutions (A,B).

\begin{figure}
\vspace{-3cm}
\plotone{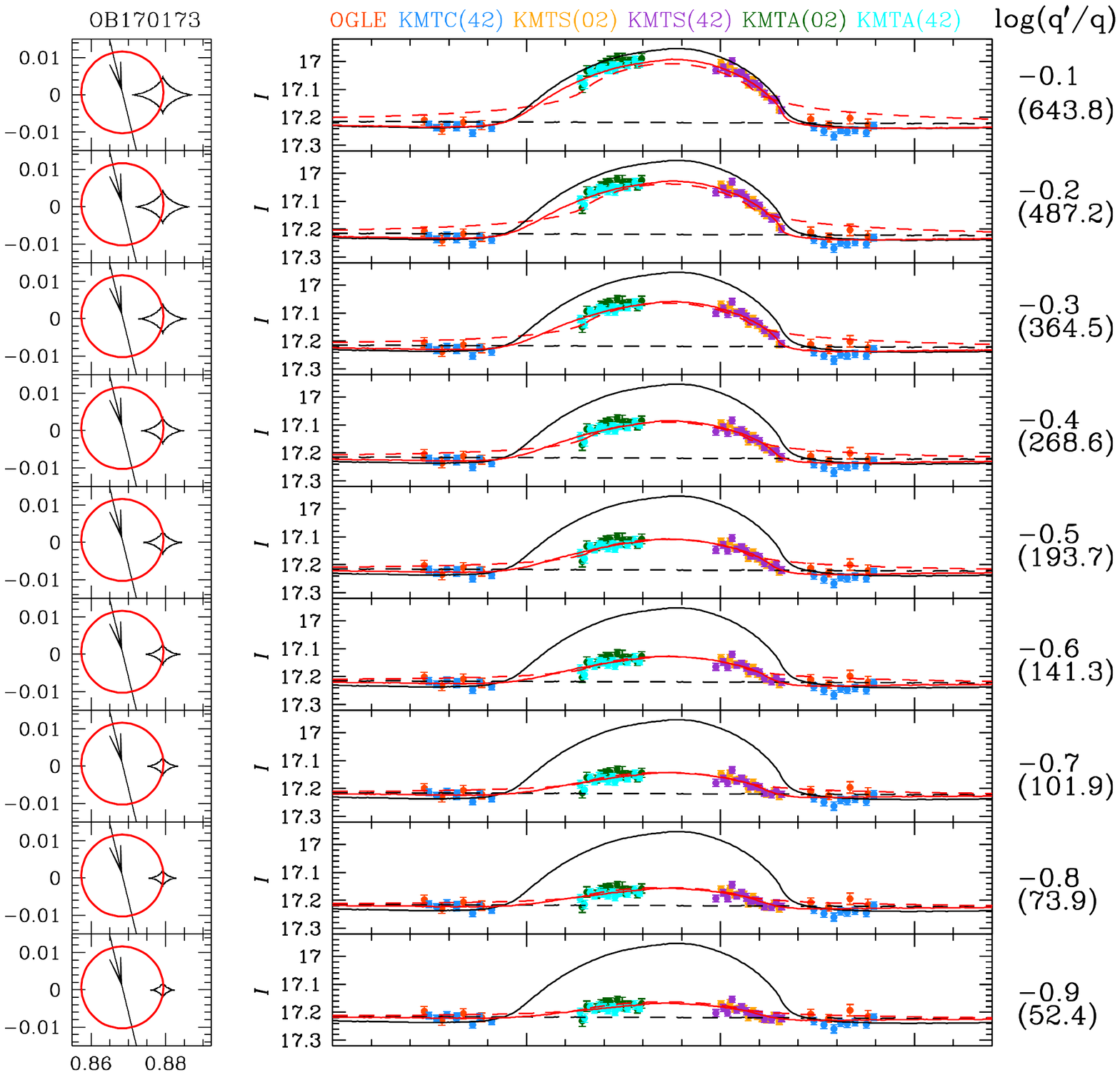}
\vspace{-3cm}
\caption{Nine simulations of OGLE-2017-BLG-0173 (von Schlieffen solution A),
similar to Figure~\ref{fig:ob171434}.   The values in parentheses
are $\Delta \chi^2 = \chi^2({\rm 1L2S}) - \chi^2({\rm 2L1S})$, by which
binary source models are excluded.  To the eye, the degeneracy between
these solutions and those in Figure~\ref{fig:ob170173b} (Cannae solution B)
persist at all $q$.
}
\label{fig:ob170173a}
\end{figure}

\begin{figure}
\vspace{-3cm}
\plotone{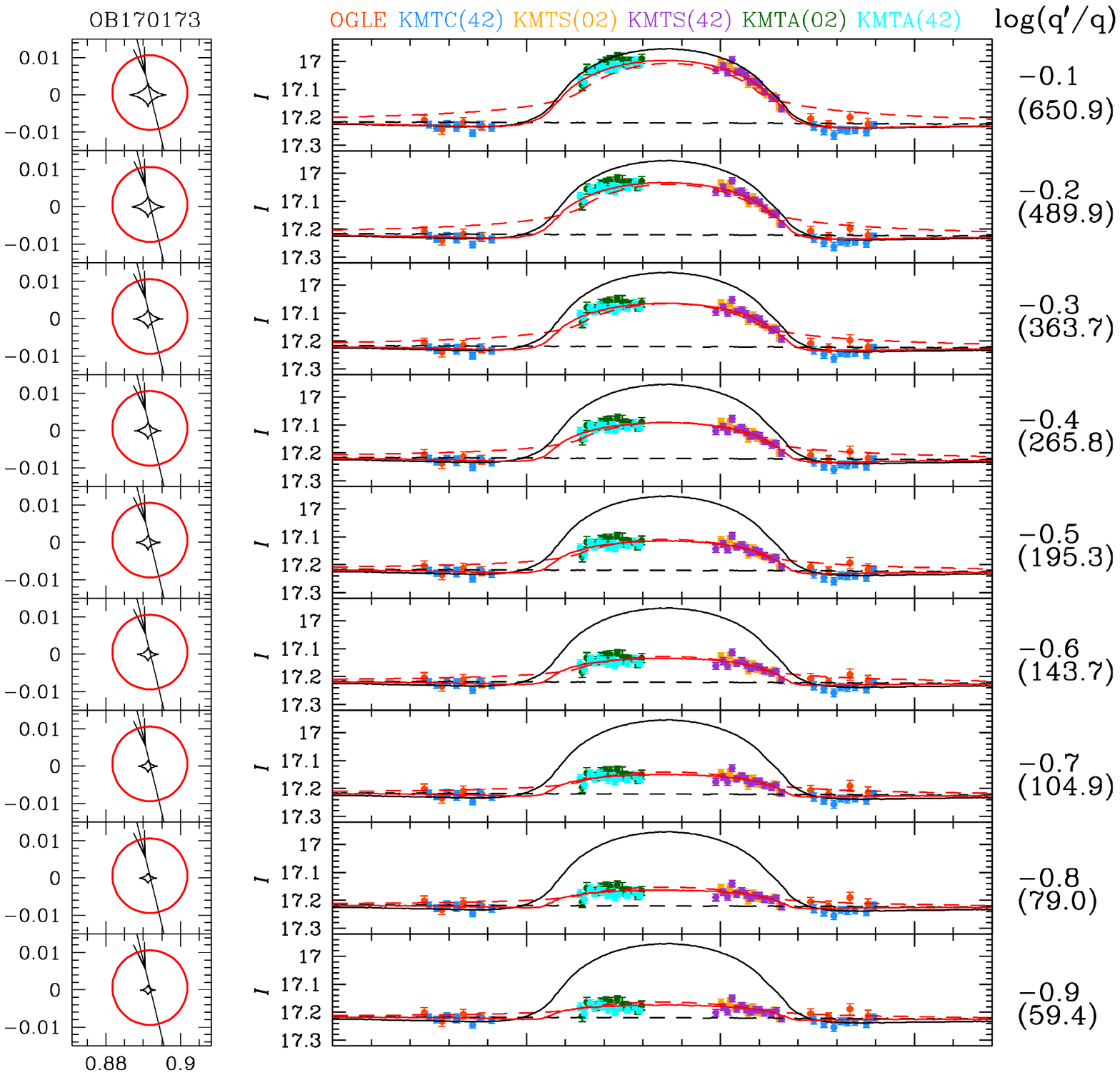}
\vspace{-3cm}
\caption{Nine simulations of OGLE-2017-BLG-0173 (Cannae solution B),
similar to Figure~\ref{fig:ob171434}.   The values in parentheses
are $\Delta \chi^2 = \chi^2({\rm 1L2S}) - \chi^2({\rm 2L1S})$, by which
binary source models are excluded.  To the eye, the degeneracy between
these solutions and those in Figure~\ref{fig:ob170173b} 
(von Schlieffen solution A) persist at all $q$.
}
\label{fig:ob170173b}
\end{figure}

As shown in their Figure~1, the event
betrays no hint of an anomaly in OGLE data, so the decision to
examine quick-look KMTNet data was not influenced in any way by the presence
of a planet.  Thus, we must evaluate the minimum value of $q$ that
would have triggered a decision to re-reduce the data and then determine
whether the resulting light curve would have been reliable enough to
warrant publication of the (putative) planet.  We note that, as shown
in their Figure~3, the ``bump'' in the KMTNet data was caused by the edge 
of a large source grazing the center of the caustic for solution (A) and
by the center of a large source passing directly over the caustic for solution
(B).
Figures~\ref{fig:ob170173a} and \ref{fig:ob170173b} show sets of 
nine simulated light curves for $\log(q^\prime/q)=-0.1,-0.2,\ldots -0.9$,
for solutions A and B, respectively.  The first point is that to 
the eye, the two figures look identical, except that the geometries
at the left are quite distinct.  Second, one sees that within each figure,
the bump looks qualitatively similar in all cases, which is fundamentally due to
the ``Hollywood'' \citep{gould97,ob170173}
character of the event.  The main difference is that
the height of the bump scales $\propto q$ (as discussed by \citealt{ob170173};
see their Equations~(9) and (10)).  We estimate that at
$\log(q^\prime/q)=-0.7$ the ``bump'' (now 0.06 mag, compared to the
actual one of 0.3 mag), would have still triggered a further investigation
for either solution A or B.  Further, the numbers at the right show
$\Delta \chi^2 = \chi^2({\rm 1L2S}) - \chi^2({\rm 2L1S})$ between
binary-source (1L2S) and binary-lens (2L1S) models.
These values are certainly high enough to exclude the 1L2S interpretation.

However, we find that at $\log(q^\prime/q)=-0.7$, and indeed at all
$q^\prime$ shown in the figures, the analysis of either simulated
data set (A or B) yields a discrete degeneracy between the two classes
of models (von Schlieffen and Cannae), with $\Delta\chi^2<10$ 
and $\Delta\log q>0.3$ between the two minima.  That is, they all
suffer from essentially the same degeneracy as the original event.
Hence, although for $\log(q^\prime/q)\geq -0.7$, we judge that they
all would have been published (just as the original event was,
\citealt{ob170173}), they all would have been excluded from the
analysis.

This exclusion has no practical importance from the perspective of
the present mass-ratio-function analysis, because the original event is itself
excluded.  However, the persistence of this degeneracy is of 
significant interest.  \citet{ob170173} had noted that the other
published Hollywood event, OGLE-2005-BLG-390 \citep{ob05390}, did not
suffer from this von Schlieffen/Cannae degeneracy.  And they further
noted that the caustic was much smaller than the source in that case,
whereas the caustic was of comparable size to the source for 
OGLE-2017-BLG-0173.  They therefore conjectured that the degeneracy
was a consequence of the caustic size relative to the source size.
However, the present analysis shows that this is clearly not the case.
Hence, there must be some other governing factor.  This may be the
angle of the source trajectory, $\alpha$, but investigation of this
question is well outside the scope of the present work.

\subsubsection{OGLE-2016-BLG-1195}
\label{sec:ob161195}
OGLE-2016-BLG-1195 was analyzed by two groups \citep{ob161195a,ob161195b}
based on completely different data sets.  The two groups obtained 
slightly different mass-ratio estimates
$q=4.22\pm 0.65\times 10^{-5}$ 
\citep{ob161195a} and $q=5.60\pm 0.75\times 10^{-5}$ \citep{ob161195b}.
Here we adopt a weighted average $q=4.81\pm 0.49\times 10^{-5}$.

The anomaly in this event was discovered and publicly
announced by the MOA collaboration in real time, i.e., at
UT 15:45 29 June.  In fact, while the internal discussions that led
to this alert were still ongoing, the MOA observers increased the cadence of
observations, beginning at UT 15:15.  That is, prior to this
change, MOA observed the field steadily at a cadence of 
$\Gamma=4.0\,{\rm hr}^{-1}$, which
is their normal cadence for this field, whereas between UT 15:15 and UT 16:48
(roughly the ``end'' of the anomaly), the field was observed 16 times,
for a mean cadence of $\langle\Gamma\rangle =10.3\,{\rm hr}^{-1}$.
A slightly lower cadence continued for the rest of the night.
See \citet{ob161195a}.

Hence, in contrast to the previous two cases, one must consider two
possible data streams for any mass ratio $q^\prime$: 
the actual one (in the case that the observers would have
recognized the anomaly and so taken the initiative to increase
the cadence) and one in which the anomaly was not recognized,
so the cadence remained at the standard $\Gamma=4.0\,{\rm hr}^{-1}$.
Fortunately, as we show below, the two cases lead to very similar
conclusions.  This is partly because the anomaly had already peaked
when the cadence was changed and partly that there existed an independent
data set from KMTA.

We first focus on the more conservative case, i.e., that at
sufficiently low $q^\prime$, the observers would not have recognized
the anomaly.  This leads us to construct a ``thinned'' version of
the MOA online data set that is consistent with the normal MOA cadence.
For the times that MOA observed the field two, three or four times consecutively
(in 1.5 min intervals), we always choose the second of these observations.
For the remaining observations, we thin them so that the surviving observations
are as closely spaced to 15 minutes
as possible.  In particular, we keep six observations from the 16 observations
during the final 93 minutes of the anomaly.

Then, under either assumption (real-time anomaly recognition or not),
and as for the two other events previously examined, 
we address two questions, whether the anomaly would ultimately have been 
recognized and,
if so, whether the re-reduced data would have yielded a reliable 
planet detection.
A very important difference from the previous two cases, however, 
is that the full extent of the anomaly was continuously observed
by two different surveys, MOA \citep{ob161195a} and KMTA \citep{ob161195b},
from sites separated by several thousand km.  The data quality was
overall roughly comparable (judged by the quoted errors of parameters).
Hence, the normal caution that a low-amplitude signal might be due
to unknown systematics would not apply to this case, since the same
signal would be present in both data sets.  Moreover, even though
the actual analyses were done in two separate papers, a joint paper
combining both data sets would have been written if neither data
set was by itself adequate for publication.  

As noted above, the first threshold is simply recognizing the anomaly.
This need not have been in real time.  In principle, it could have
been recognized later in either the MOA or KMTNet data.  However, in contrast
to the two cases discussed above, the 2016 quick-look KMTNet
data were of substantially lower quality, and tests show that
the anomaly would not have been recognized according to the procedures
followed in that year.  Hence we should examine only the on-line
MOA data.  We note that the event was a {\it Spitzer} microlensing
target \citep{ob161195b}, part of a broader program to measure the
Galactic distribution of planets 
\citep{prop2015a,prop2015b,yee15,21event,zhu17},
and therefore the MOA data would have been examined quite closely even
if the anomaly had not been detected in real time.

Following the above considerations, we examined fake light curves
constructed on the basis of MOA online data, both with and without
the additional MOA points triggered by the alert.  We are confident
that the anomaly would have triggered re-reductions of MOA and KMTNet
data for $\log(q^\prime/q)=-0.3$ and perhaps even lower.  However,
we do not investigate the exact threshold, nor do we show the plots that
we reviewed because, as we now describe, the fundamental issue is
not simply recognizing that there was an anomaly.

\begin{figure}
\vspace{-3cm}
\plotone{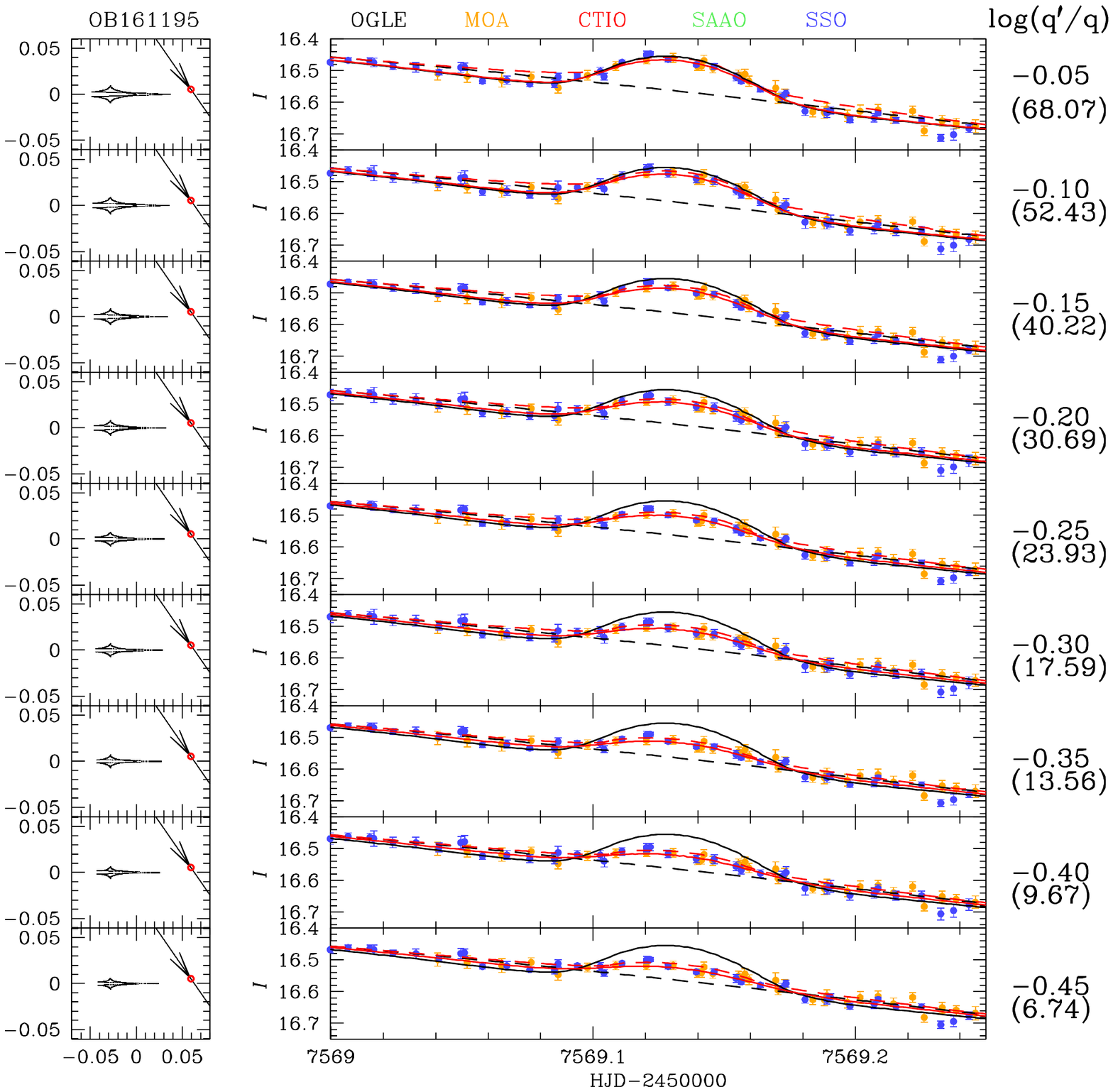}
\vspace{-3cm}
\caption{Nine simulations of OGLE-2016-BLG-1195, similar to 
Figure~\ref{fig:ob171434}, 
except that in this case the data
points are based on the re-reduced data in order to focus on whether
the event (once recognized as interesting) would be publishable.
}
\label{fig:ob161195}
\end{figure}

Figure~\ref{fig:ob161195} 
shows 9 panels with $\log(q^\prime/q)=-0.05,-0.10,\ldots -0.45$,
with re-reduced data from both MOA and KMTA.  In each panel, we show
both the planetary (2L1S) and binary source (1L2S) models.  The number
in parentheses to the right of each panel gives the
$\Delta\chi^2\equiv\chi^2({\rm 1L2S})-\chi^2({\rm 2L1S})$ difference between
these models.  For $\log(q^\prime/q)=-0.30$, $-0.35$, and  $-0.40$,
these are 17.6, 13.6, and 9.7 respectively.  Given that two independent
data sets are contributing to these values and that neither shows
any sign of systematics, we conclude that the first two would be considered
adequate for a reliable detection of a planet, while the third would not.

Next, we address the role of the real time alert.  If the event
were not recognized in real time at lower mass ratio (as it was in
the actual case) then the only difference in the evaluation
of publishability would be the ``exclusion'' of $\sim 10$ MOA points 
during the second half of the anomaly (plus some post-anomaly points), which
would imply slightly lower $\Delta\chi^2$.
In particular, for $\log(q^\prime/q)=-0.30$ and $-0.35$, $\Delta\chi^2=13.1$
and 10.2, respectively.  By the argument just given, these would render
the first as a publishable planet and the second not.  Hence, the
only relevant question about real-time alerts is whether, at 
$\log(q^\prime/q)=-0.35$, the event
would have been alerted based on the online MOA data.
Given that the actual event was not recognized by the observers 
until the deviation was nearly at peak, and so with roughly twice the deviation
as the $\log(q^\prime/q)=-0.35$ case, we regard this as highly unlikely.
Thus, we conclude that the threshold for detection/publication is
$\log(q^\prime/q)=-0.30$.

Finally, we analyze the simulated $\log(q^\prime/q)=-0.30$ as though it were
a real event.  We find that there is a unique minimum and that
$\sigma(\log(q))=0.10$.  Hence, it clearly meets our criterion (2).

\subsubsection{OGLE-2013-BLG-0341}
\label{sec:ob130341}

OGLE-2013-BLG-0341 was originally recognized as
having a minor-image planetary anomaly at HJD$^\prime=6393.7$ based
on real-time OGLE data, four days after the anomaly, when A.G. cataloged
it as a possible high magnification event and therefore carefully examined
the light curve as part of his assessment.  He publicly announced
this but also noted that since the lower limit on the magnification,
$A_\max>10$, was not particularly high, no immediate action was
warranted.  However, from this point on, the event was closely monitored
with the aim of organizing intensive followup observations near peak,
if indeed its prospective high-mag character was confirmed.  Such intensive
observations were in fact initiated 10 days after the anomaly
and about two days before peak.  The peak was characterized primarily
by a binary (not planetary) caustic.  However, \citet{ob130341} later
showed that the planet's parameters could be recovered even when
the data points in the vicinity of the planetary anomaly 
(144 points. to be precise) were removed from the data.

The initial recognition of the planetary dip can be partially attributed
to chance.  The purpose of A.G.'s review of the online OGLE light curves
was not to find planetary anomalies, but rather high-magnification events
that could be intensively monitored to find planets near peak \citep{gould10}.
Thus, it was hardly guaranteed that the actual signal would have been 
noticed when it was, and it is fairly unlikely that it would have been 
noticed at this point if the dip had been only half as 
deep.

Nevertheless, because the event eventually did become high-magnification,
the light curve would have been singled out for extremely close
inspection, regardless of whether a planetary anomaly had previously
been noticed or not.  If the planet had been noticed at this point, then
of course exactly the same followup observations would have been initiated.
If not, then when the event showed very obvious binary-like behavior
(see Figure~1 of \citealt{ob130341}), the event would have been
abandoned as ``not interesting''.  Hence, the
signature from the planet in the caustic exit would have been drastically
reduced due to lack of follow-up data\footnote{
With respect to the coverage of the caustic, both the OGLE and MOA 
``survey'' data must be treated  substantially as ``followup''.
For the night of the caustic, MOA increased its cadence from its normal
level of $\Gamma=3.5\,{\rm hr}^{-1}$ to $\Gamma=20\,{\rm hr}^{-1}$.
For OGLE, the role of the alert was somewhat more involved.
OGLE-2013-BLG-0341 lies close to the edge of the OGLE chip BLG501.12.  
Hence, it tended to drift off toward -- and then over -- the edge of the chip
over the course of a given night, until the telescope pointing was
reset.  Thus, for example, the sixth and final point on the night of 
the ``dip'' was at ${\rm HJD}^\prime=6393.717$, whereas other events
in this field had nine additional points on this night ending at
${\rm HJD}^\prime=6393.909$.  The pointing for this field was specially
adjusted on the night of the caustic exit, which probably 
roughly doubled the number of OGLE points that night relative to what would
have occurred in the absence of an alert.  Hence, the alert accounted
for not only 100\% of the data from non-survey telescopes, but also
82\% of the MOA data and 50\% of the OGLE data.  Only the Wise survey
was unaffected by the alert.}.

In brief, an important (though as we shall see, not only) 
question of whether the planet would have been
recognized comes down to:  would \citet{ob130341} have recognized
the planetary signature once they began closely examining this
high-mag event, as it approached peak?

\begin{figure}
\vspace{-3cm}
\plotone{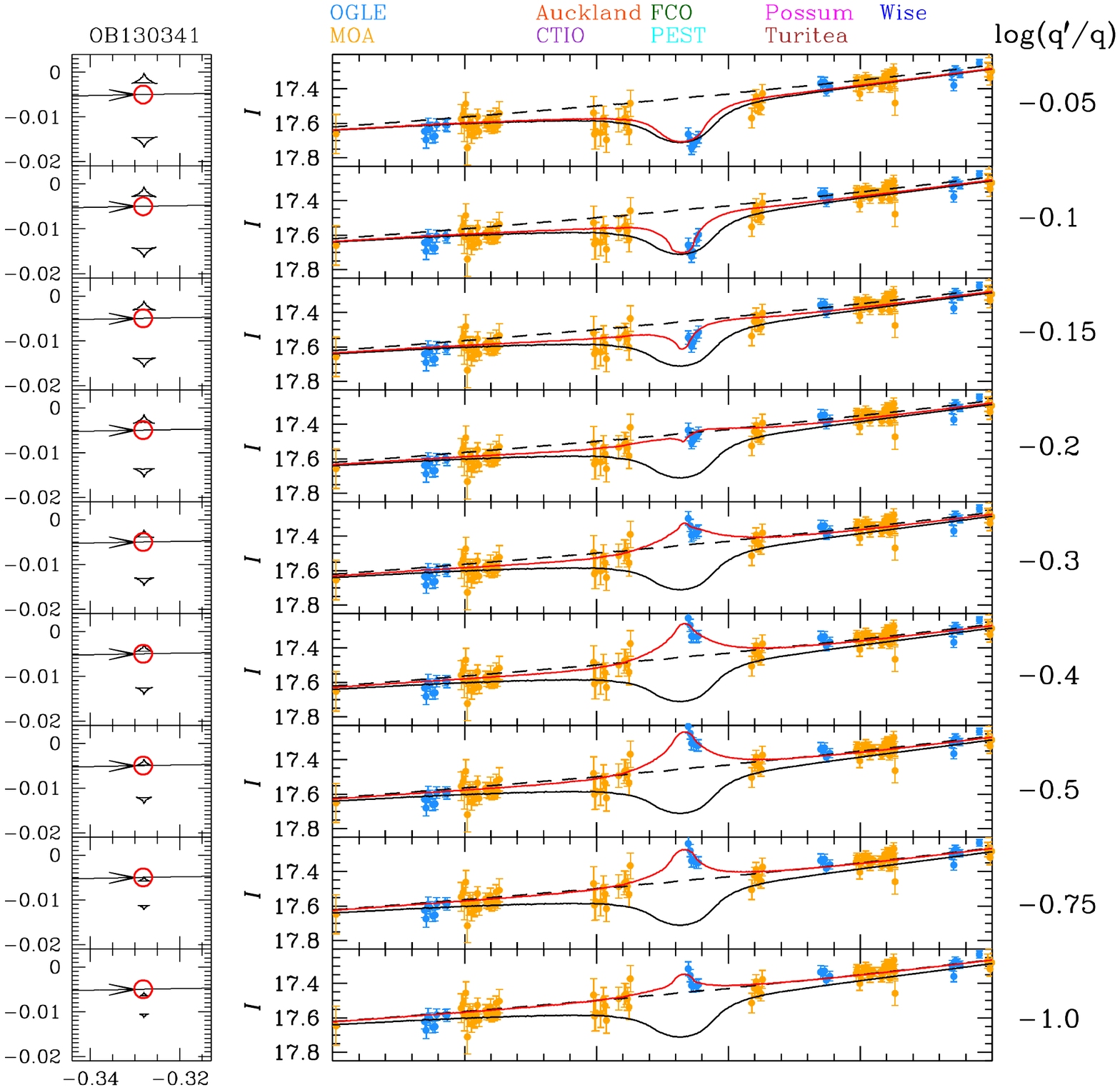}
\vspace{-3cm}
\caption{Nine simulations of OGLE-2013-BLG-0341, similar to 
Figure~\ref{fig:ob171434}.  Similar to those simulations, it is
based on ``online'' OGLE and MOA data in order to focus on the
problem of real-time recognition of the planetary anomaly.
}
\label{fig:ob130341}
\end{figure}

Figure~\ref{fig:ob130341} shows nine panels with\hfil\break\noindent
$\log(q^\prime/q)=-0.05,-0.10,-0.15,-0.2,-0.3,-0.4,-0.5,-0.75,-1$.
This event is unique in our sample in that as $q$ declines, the anomaly
becomes less visible and then invisible, but then begins to become
more visible again at $\log(q^\prime/q)=-0.3$.  
Then, at $\log(q^\prime/q)=-0.5$, its visibility peaks, whereupon it
gradually declines.  The reason for this behavior is that in the actual
event, the source passed along the trough between the two triangular,
minor-image caustics, but very close to one of the caustic walls.
The source trajectory relative to the mid-line of the trough remains
basically the same as $q$ declines, but because the triangular caustics
move closer together, the edge of the source moves
increasingly over the caustic wall.  At first, the excess magnification
of the limb increasingly cancels the dip due to the main part of the
source passing through the trough.  However, when $q$ falls sufficiently,
the excess magnification starts to dominate, and there is a bump in place
of the trough.

The scatter in the OGLE online data for the six points
on the night of the anomaly is small, $\sigma=0.03\,$mag.  Hence,
a mean deviation of just 0.06 mag would be a five-sigma detection,
which we regard as the minimum needed to be both recognizable by
eye and to engender sufficient confidence to trigger a massive
followup campaign, as discussed above.  This threshold has already
been crossed at
$\log(q^\prime/q)=-0.15$ for the ``fading'' dip.  It is crossed again
for the ``rising'' bump at $\log(q^\prime/q)=-0.3$ and then is
crossed again for the ``fading'' bump slightly below
$\log(q^\prime/q)=-0.75$.  We conclude that followup observations
would have been triggered in the two ranges 
$(\log(q^\prime/q)>-0.1)$ and $(-0.3>\log(q^\prime/q)>-0.75)$.

Nevertheless, we now argue that only in the former range would a paper
claiming secure detection of a planet have been written.  First, in contrast
to a dip in the light curve (which can only be explained by a minor-image
anomaly), an isolated bump in the light curve can also be explained
by a 1L2S solution.  The OGLE anomaly data are confined to a narrow
range in time, and so have no leverage on the shape of the bump to
distinguish between 2L1S and 1L2S.

In principle, the followup data (which we argued above would have been
triggered for the second -- ``bump'' -- range of $q$) could have
confirmed the planetary nature of the anomaly.  However, there are two
practical issues that severely undermine this possibility.  First,
at our finally adopted value of $\log(q^\prime/q)\leq -0.1$, 
we find that this confirmation is already
relatively weak, $\Delta\chi^2=66$, a point to which we return below.
More fundamentally, it is very unlikely that the analysis would have been
pushed to the point that the test would have been devised of
deleting the planetary-anomaly data.  The analysis of the event
was quite complex and required enormous human and computer resources.
It was only in the process of carrying out this analysis that it was
discovered that such confirmation was possible.  Hence, the motivation
to analyze an event to this level in the case that there was a
(seemingly) unassailable non-planetary interpretation would almost
certainly have been lacking.  Finally, even supposing that such
an analysis were done, the ``confirmation'' of $\Delta\chi^2<50$
would almost certainly not have been regarded as sufficient to claim
a secure detection.  This is reflected in the fact that 
\citet{ob130341} specifically argued that the unambiguous
planetary anomaly (due to the fact that it was a minor-image dip)
served as ``confirmation'' that the subtle -- and by eye, invisible --
deviations in the binary caustic could be regarded as a reliable
indicator of a planet.  Without such independent knowledge, and
at relatively low $\Delta\chi^2$, this would have at best led to
the reporting of an interesting planetary candidate.

We conclude that the threshold for planet detection is $\log(q^\prime/q)=-0.10$.
We confirm that this solution is unique and find $\sigma(\log(q))=0.04$,
implying that our sample criteria are satisfied at this 
threshold\footnote{We note that OGLE-2013-BLG-0341 was part of the 
\citet{shvartzvald16} sample of events that were jointly monitored by
the OGLE, MOA, and Wise surveys.  All such events were examined
extremely closely by Y.S., so that it is possible that the planetary
anomaly would have been noticed after the event was over, even if it were
missed prior to the peak.  In this case, however, there would have
been no follow-up data and therefore only extremely weak confirmation of
the planet from the binary caustic.}.

\subsubsection{MOA-2009-BLG-266}
\label{sec:mb09266}

MOA-2009-BLG-266 \citep{mb09266} was recognized as having
a potentially planetary anomaly in real time by the MOA collaboration
at the end of the New Zealand night from the sharp decline in the
previously smoothly rising light curve.  This triggered
follow up observations at many sites,
which further articulated the decline,
mapped the trough and then the rise.  
The basic model of the
event was already derived before any of these followup data were taken,
let alone reduced, so that in the actual case, the alert-generated
data were needed for full characterization of the planet, but not
for its discovery.  However, if the mass-ratio had been lower, it 
is possible that the planet could not have been characterized well
enough to warrant publication in the absence of the followup data.
Thus, for this event, it is especially important to evaluate both
how well the planetary perturbation could have been recognized in
real time, and how well the planet could have been characterized
with, and without, followup observations.

We note that five different observatories took data on MOA-2009-BLG-266
prior to the alert, of which four also took data after the alert.
Hence, one must assess whether these four would have taken data
in the absence of the alert.  We find that only one of these
(Canopus) took data in a way that indicated sustained focus on the
event as it approached peak: they took four points spread over 1.4
hours on the night before the alert.  The others either took
one point on occasional nights or had stopped taking data altogether.
Thus, it is reasonable to suppose that Canopus would have also taken
four points on the next night, even if there had been no alert.
However, these data would have overlapped MOA data and so would not
have qualitatively altered how well the event could have been
characterized in the absence of an alert (and so absence of data in the trough).

\begin{figure}
\vspace{-3cm}
\plotone{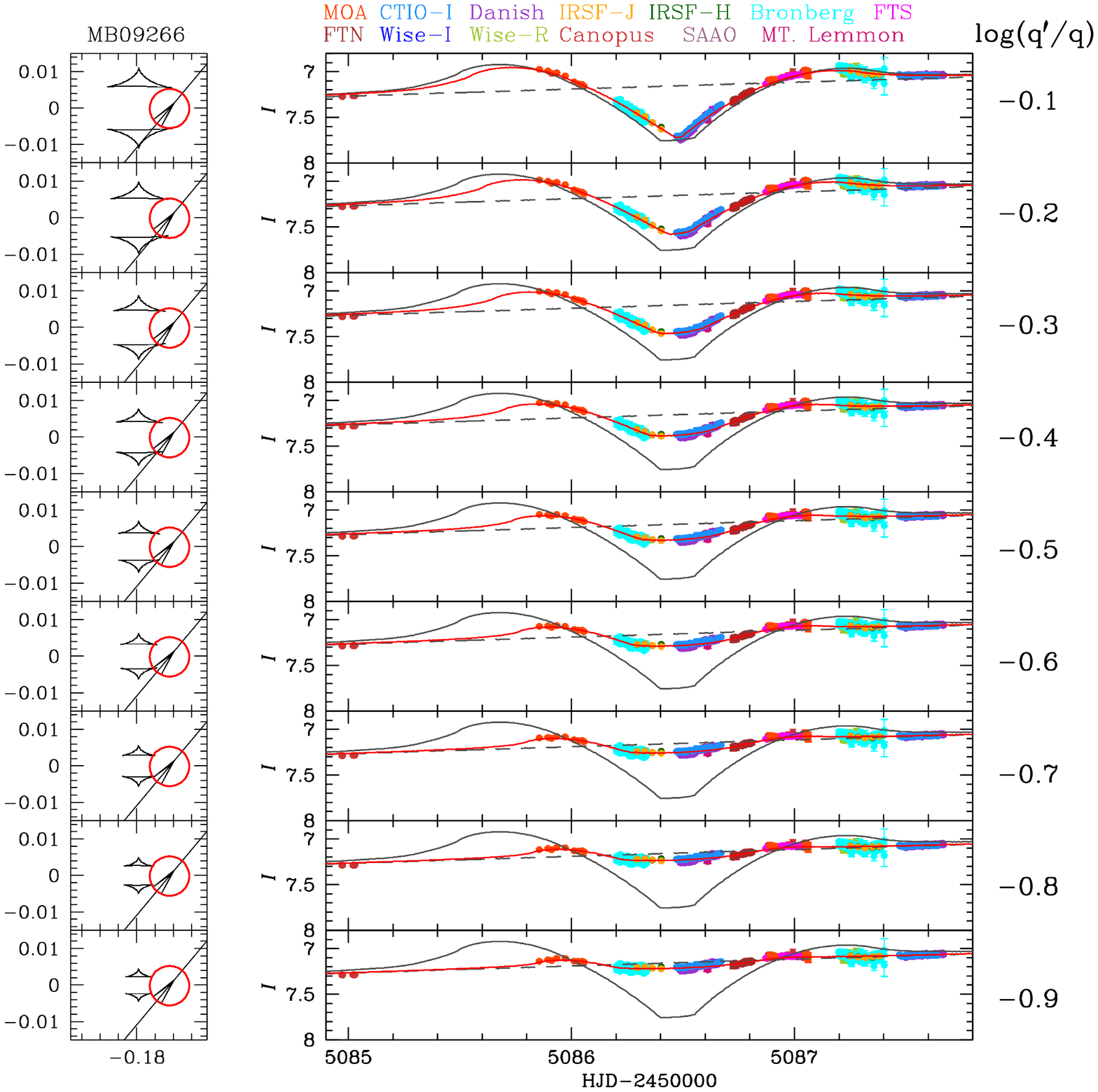}
\vspace{-3cm}
\caption{Nine simulations of MOA-2009-BLG-266, similar to 
Figure~\ref{fig:ob171434}.  The simulations are based on re-reduced
data from all observatories.
}
\label{fig:mb09266}
\end{figure}

Figure~\ref{fig:mb09266} shows nine panels with
$\log(q^\prime/q)=-0.1,-0.2,\ldots -0.9$ and all data re-reduced.
The first question is whether, with a smaller mass ratio, MOA
would have issued an alert (based, of course, on online data).
While Figure~\ref{fig:mb09266} shows re-reduced data, it still enables us to
understand how the basic form of the MOA light curve evolves as
$q$ declines: over the range $\log(q^\prime/q)\leq -0.6$, 
it basically takes the form of a mean excess over the 
point-lens model (dashed line).  We now argue that an anomaly of this
form would give rise to an alert provided that the mean excess over the
predicted point-lens light curve was 0.1 mag.  

Based on the MOA online data from before the night of the anomaly,
we find that one can predict the flux (based on a point-lens model)
on the night of the anomaly to 0.085 mag at $3\,\sigma$ confidence.
There are 10 MOA points on that night, with scatter 0.024 mag.   Hence
a standard error of the mean of 0.008 mag. Thus, a mean offset
of 0.1 mag would yield a $\Delta\chi^2\sim 12$ discrepancy, which
would be sufficient indication to issue an alert.  This condition is
satisfied for $\log(q^\prime/q)= -0.6$, but not lower mass ratios.
At higher mass ratios, the mean offset itself satisfies this condition,
and in addition there is increasing evidence of a decline (which
is what triggered the actual alert).

We fit simulated data for the case $\log(q^\prime/q)= -0.6$, and find that
the solution is both unique and well localized ($\sigma(log(q))=0.02$).
We then repeat this
exercise for $\log(q^\prime/q)= -0.7$ but with only MOA and Canopus data
(since, as we argued above, there would be no alert and hence no
followup data apart from Canopus).  We find that there are several local
minima from the broad search of parameter space, and that none of the
models derived at these minima appear compelling enough to
warrant publication.

We conclude that the threshold for this event is
$\log(q^\prime/q)= -0.6$.

\subsubsection{OGLE-2007-BLG-368}
\label{sec:ob07368}

The details of the anomaly alert of OGLE-2007-BLG-368 are recounted
by \citet{ob07368}.  The first alert was given by the robotic SIGNALMEN
anomaly detector \citep{signalmen} JD$^\prime = 4302.314$, being
triggered by the nine MOA points that lie $\sim 0.2\,$mag below
the point-lens model.  This alert
prompted followup observations beginning 5 hours later in Chile
by the $\mu$FUN SMARTS 1.3m telescope and the PLANET Danish 1.5m,
and then from additional telescopes continuing toward the west.  From the
present perspective, it is important to note that this alert did
not reach the MOA observer and so did not influence the cadence
of MOA observations on the night HJD$^\prime \sim 4303$.  These
were the next observations after those in Chile.  Hence, the next
observations that were influenced by the alert (after Chile) were
from the PLANET Canopus observations from Tasmania, whose
three closely spaced points basically overlap the last MOA point.
There were additional followup observations, which played an 
important role in characterizing the actual event, but as we will
show, these play very little role in the current $V/V_\max$ analysis.

\begin{figure}
\vspace{-3cm}
\plotone{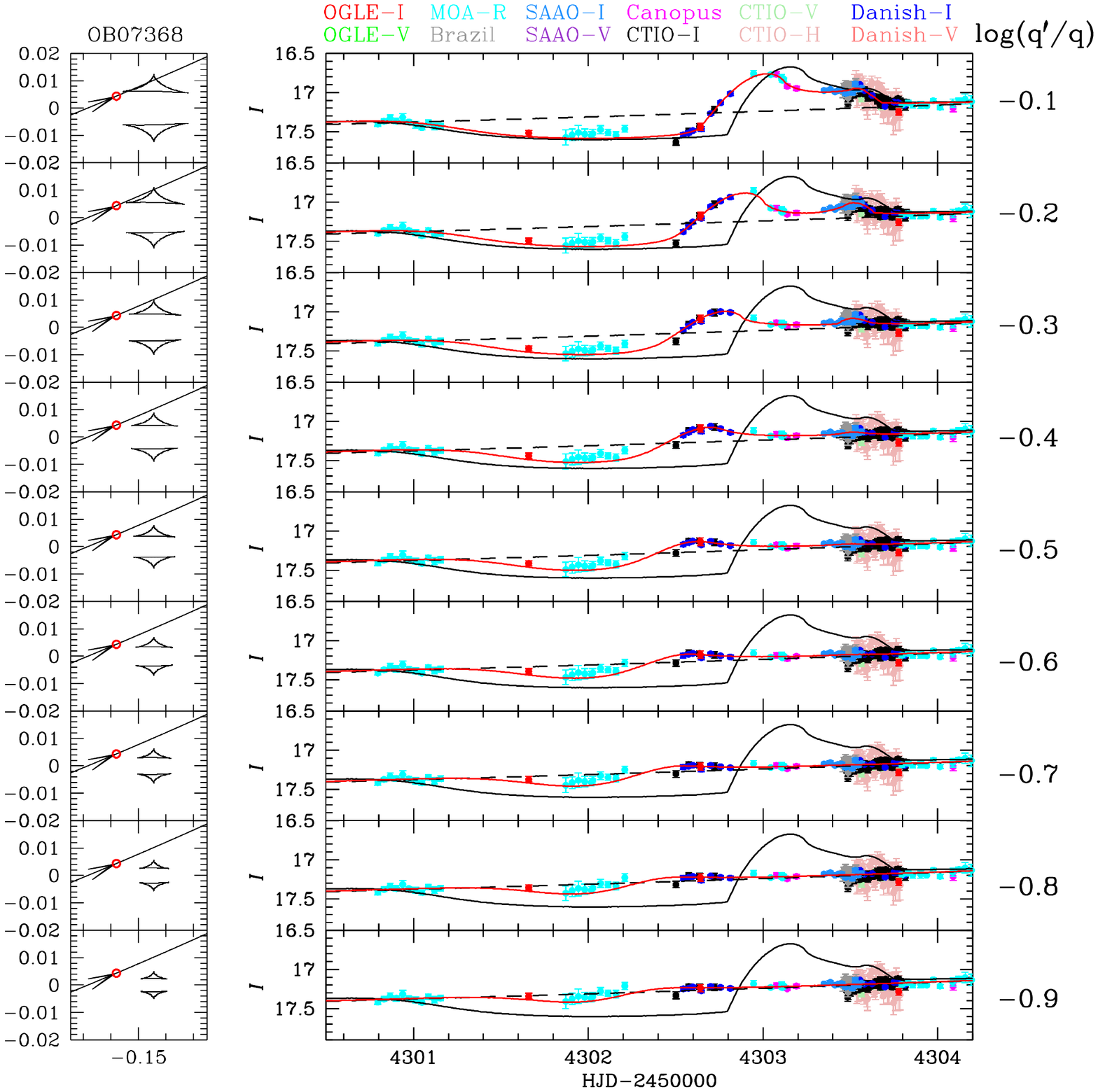}
\vspace{-3cm}
\caption{Nine simulations of OGLE-2007-BLG-368, similar to 
Figure~\ref{fig:ob171434}.  This figure is based on re-reduced
data from all observatories.  It should be compared to  
the next one (Figure~\ref{fig:ob07368online}), which 
is based on ``online'' OGLE and MOA data.
}
\label{fig:ob07368}
\end{figure}

Figure~\ref{fig:ob07368} shows nine panels with
$\log(q^\prime/q)=-0.1,-0.2,\ldots -0.9$ and all data re-reduced.
Note that for $\log(q^\prime/q)\leq-0.4$, the point-lens model and
the planetary model are nearly identical for the followup data taken
HJD$^\prime >4303.3$.  That is, only the $\mu$FUN Chile, Danish Chile, 
and Canopus data would have played a significant role for 
$\log(q^\prime/q)\leq-0.4$.

\begin{figure}
\vspace{-3cm}
\plotone{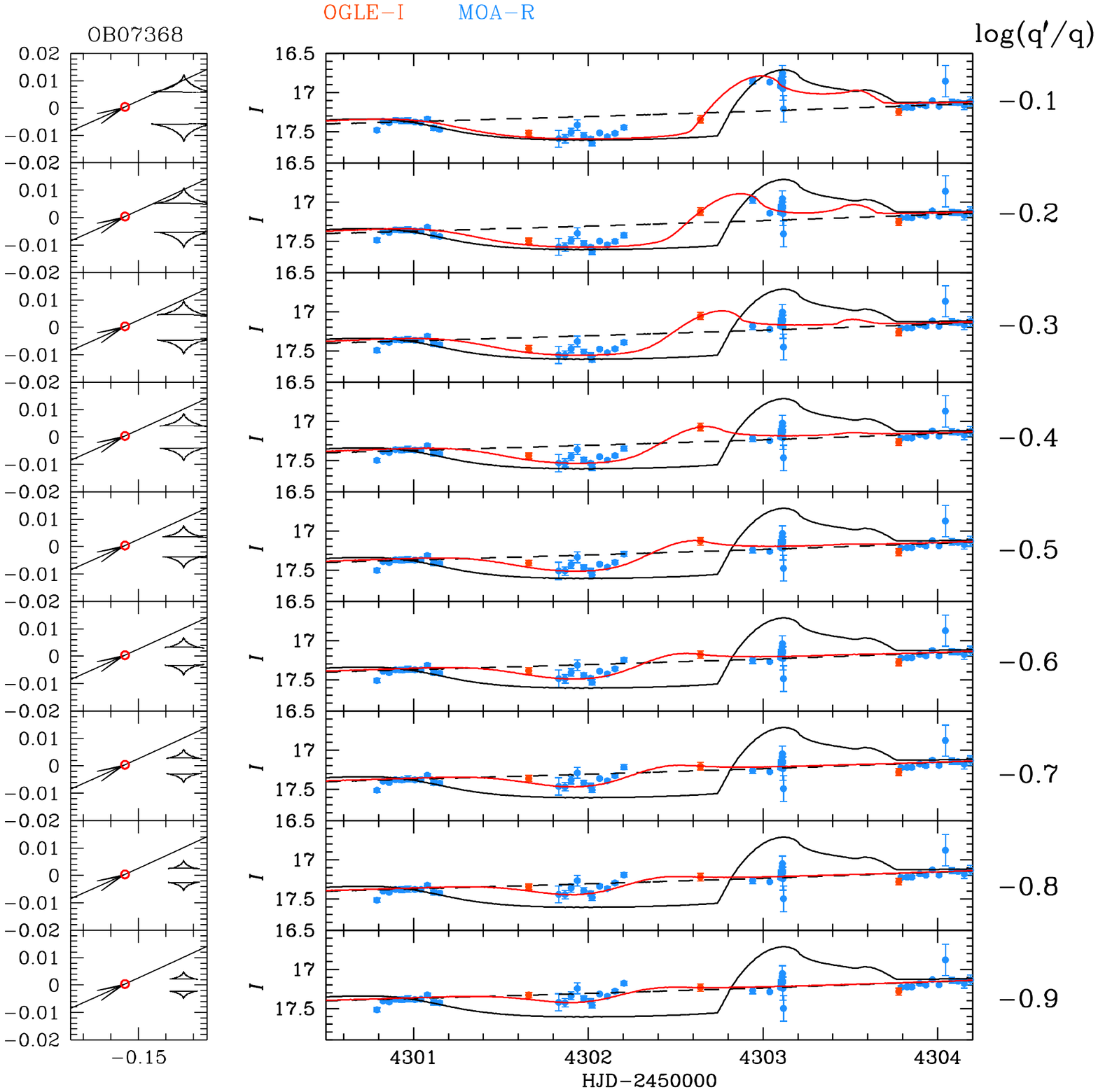}
\vspace{-3cm}
\caption{Nine simulations of OGLE-2007-BLG-368, similar to 
Figure~\ref{fig:ob171434}.  This figure is based on ``online'' 
OGLE and MOA data. It should be compared to  
the previous one (Figure~\ref{fig:ob07368}), which 
is based on re-reduced data from all observatories.  
}
\label{fig:ob07368online}
\end{figure}

Figure~\ref{fig:ob07368online} 
shows the same nine panels as Figure~\ref{fig:ob07368} but with only online
survey data.  Based on this figure, we consider it to be unlikely
that there would have been an alert on this event in time to trigger
CTIO observations for
$\log(q^\prime/q)\leq -0.4$.  Moreover, we can say with near certainty
(since A.G.\ made this decision) that CTIO would not have responded to
such an alert if it had been given.  However, the CTIO response is of secondary
importance because the Danish data, which cover the same time interval,
would certainly have been taken.

As usual, we first ask at what threshold
would the online survey data have led to re-reductions, and
then ask whether these reductions would have led to a publishable
result given the data that would have been acquired.

The online OGLE data would, by themselves, certainly have triggered 
re-reductions at $\log(q^\prime/q)\geq -0.4$.  At $\log(q^\prime/q)= -0.5$
this is less probable, but in this case the partial corroboration from
online MOA data would have almost certainly led to re-reductions.
Re-reduction at $\log(q^\prime/q)= -0.6$ is also a possibility.

Recall that at $\log(q^\prime/q)= -0.3$, we concluded that there would
have been an anomaly alert.  We analyze all data and find that the
solution is well-localized with $\sigma(\log q)=0.025$.  
At $\log(q^\prime/q)= -0.4$, we must consider two cases, i.e., with
and without the alert (and so followup data).  We find that with the
followup data, the minimum is well localized with $\sigma(\log q)=0.08$,
so satisfying all our criteria.  However, with survey-only (re-reduced)
data, we find that there are two minima separated
by $\Delta\chi^2<10$ and $\Delta\log q>0.3$, which would fail our
third criterion.  Since we assessed that there would probably not
be an alert at $\log(q^\prime/q)= -0.4$, we conclude that the threshold 
for detection is $\log(q^\prime/q)= -0.3$.  We recognize that there is
some probability of an alert at $\log(q^\prime/q)= -0.4$, and therefore
(according to the analysis just given) of a publishable detection.
However, since we are approximating probabilities $P$ as either zero
or one, we simply adopt the threshold $\log(q^\prime/q)= -0.3$.

\subsubsection{OGLE-2005-BLG-390}
\label{sec:ob05390}

OGLE-2005-BLG-390 was detected primarily in follow-up data organized
by the PLANET collaboration, but in contrast to all previous cases,
all of these data were taken in response to an alert of the microlensing
event itself, not an anomaly \citep{ob05390}.  The anomalous behavior
was noted by the observer at the Danish telescope in Chile, and
in principle this could have influenced other observatories farther
to the west, but a detailed investigation at the time showed that
this was not the case.  Hence, the same observations would have been
taken over the anomaly whether a planetary signal was suspected to
be present or not.

\begin{figure}
\vspace{-3cm}
\plotone{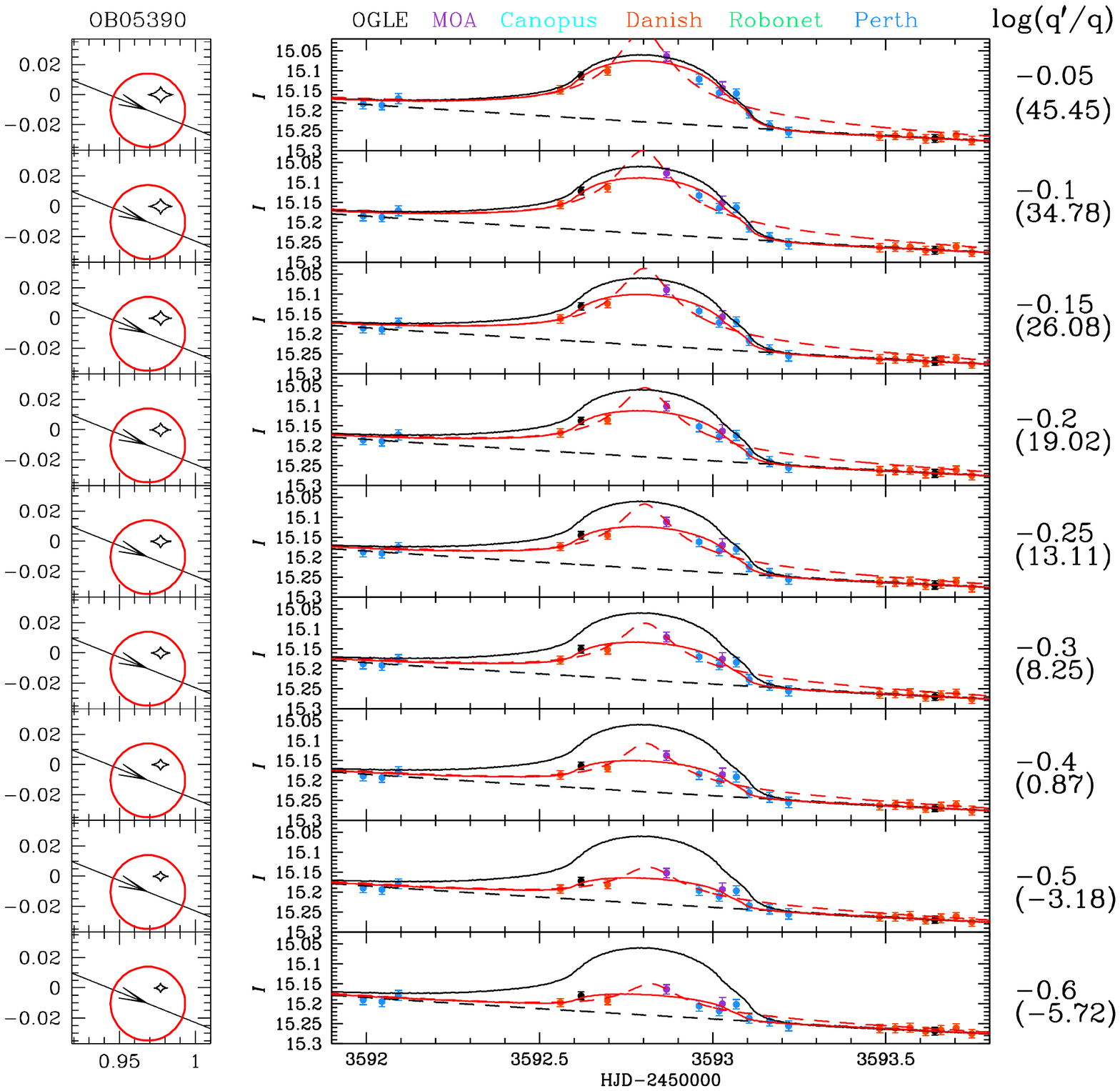}
\vspace{-3cm}
\caption{Nine simulations of OGLE-2005-BLG-390, similar to 
Figure~\ref{fig:ob171434}. It is based on
re-reduced data from all observatories, since these reductions would
have been carried out whether or not a planet was suspected.
}
\label{fig:ob05390}
\end{figure}

Figure~\ref{fig:ob05390} shows nine panels with
$\log(q^\prime/q)=-0.05,-0.10, \ldots ,-0.30,-0.40,-0.50,-0.60$ 
and all data re-reduced.
This event was one of relatively few monitored by the PLANET collaboration,
and hence the data would have been very closely inspected even if the anomaly
had not been noticed in real time.  Thus, the anomaly would have been easily 
noticed at 
$\log(q^\prime/q)=-0.5$ and probably somewhat below.  However, as we now
show, it is not important to evaluate the exact boundary of this
recognition.

Because the anomaly is a smooth bump, it can potentially be fit
as a binary source event, 1L2S.  In the actual event, this degeneracy
was investigated and ruled out both by $\Delta\chi^2=46$ and also by
obvious deviations in the light curve from the 1L2S model.  However,
for $\log(q^\prime/q)=-0.25$, we see that 2L1S is favored over 1L2S
by only $\Delta\chi^2=13$.  For the case of OGLE-2016-BLG-1195, we
regarded this value as marginally acceptable because there were
two independent data sets that spanned the entire anomaly.  However,
in the present case, which lacks such confirmation, we regard
it to be marginally unacceptable and therefore adopt 
a threshold of $\log(q^\prime/q)=-0.2$.  We find that at this
threshold, $\sigma(\log q)=0.11$, which satisfies our sample criterion
$\sigma(\log q)< 0.15$.

\subsubsection{OGLE-2005-BLG-169}
\label{sec:ob05169}

\begin{figure}
\vspace{-3cm}
\plotone{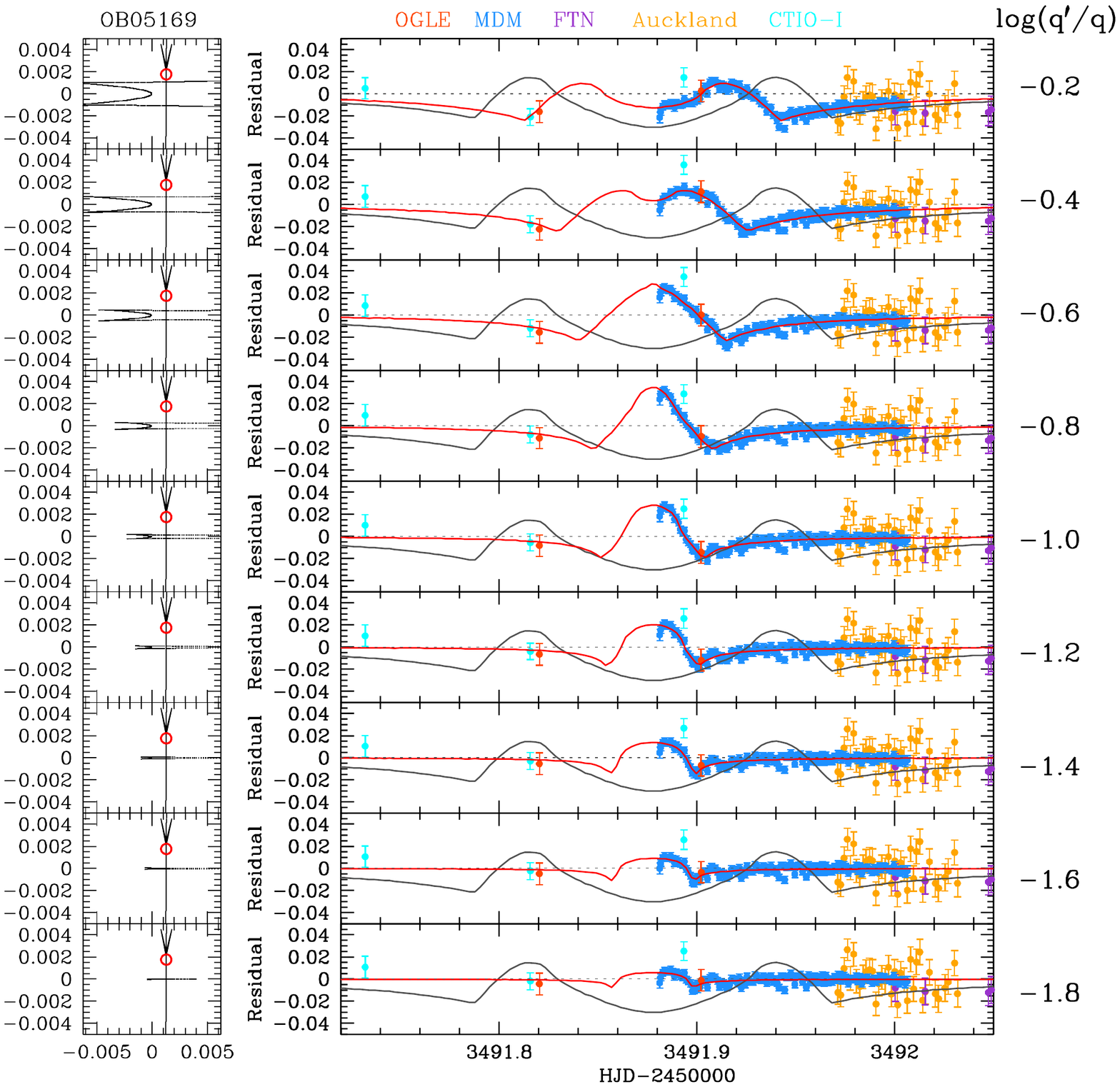}
\vspace{-3cm}
\caption{Nine simulations of OGLE-2005-BLG-169, similar to 
Figure~\ref{fig:ob171434}. It is based on
re-reduced data from all observatories, since these reductions would
have been carried out whether or not a planet was suspected.
}
\label{fig:ob05169}
\end{figure}

The overwhelming majority of evidence for a planet in the OGLE-2005-BLG-169
data lies in the extremely intensive followup data taken from the MDM
observatory in Arizona, beginning very close to the peak of the event
\citep{ob05169}.  As with all data in and near the anomaly of 
this event (and like OGLE-2005-BLG-390), these were obtained without
reference to the possible existence of a planet.  Because the dense
data began near peak, there was a modest ambiguity in the solutions
presented by \citet{ob05169}. This was resolved by the analyses of
\citet{ob05169ben} and \citet{ob05169bat} when they, respectively,
partially and fully resolved the lens 6.5 and 8.2 years after the event.
Here we use their parameters for the event, in particular 
$q=6.1\times 10^{-5}$.
Figure~\ref{fig:ob05169} shows nine panels with
$\log(q^\prime/q)=-0.2,-0.4,\ldots -1.8$, with all data re-reduced.

Because of the unprecedented character of the data set, roughly 1000
high-precision data points (later binned to 137)
taken over three hours, the data were examined
extremely closely.  Although the deviation from a point lens was
noticed very quickly, submission of the manuscript was delayed
for 10 months, primarily because of concerns that the quite small
amplitude of the deviations might be due to variable weather conditions,
which were very severe during the night of the anomaly.  In the end,
the decision to publish was based on the unambiguous discontinuous
change of slope at HJD$^\prime=3491.97$.  Such discontinuities are
a generic feature of microlensing caustic crossings but would be extremely
difficult to produce by weather-induced artifacts.

Using the same criteria (and relying on the judgment of A.G., who made
the original decision to publish) we conclude that the simulation with
$\log(q^\prime/q)=-1.2$ would marginally meet this condition.  
We fit simulated data at this value and find that 
$\sigma(\log q)=0.174$, which does not satisfy our sample criterion.
However, at $\log(q^\prime/q)=-1.0$, we find $\sigma(\log q)=0.10$, and
so adopt $\log(q^\prime/q)=-1.0$ as our threshold.

\subsection{Constraints of the Mass-Ratio Function $F(q)$}
\label{sec:foqconstraint}

If $F(q)$ is chosen correctly, then we expect the seven $r_i$ defined
by Equation~(\ref{eqn:vvmax}) to be uniformly distributed on the
interval [0,1].  As discussed in the separate analyses of each event
in Section~\ref{sec:Pqeval} (and in particular, in 
Section~\ref{sec:ob130341}), in all cases $P_i$ takes the form
$P_i(q') = \Theta(10^{-4}-q)\Theta(q-q_{\min,i})$, where $\Theta$
is a Heaviside step function and $q_{\min,i}$ has been evaluated separately
for each event.  Hence Equation~(\ref{eqn:vvmax}) reduces to 
Equation~(\ref{eqn:vvmaxspec}) 

We expect then that
\begin{equation}
{1\over N}\sum_{i=1}^N r_i = {1\over 2}\pm(12 N)^{-1/2}\longrightarrow
0.500\pm 0.109,
\label{eqn:rbar}
\end{equation}
where $N=7$.  We also expect that the distribution of $r_i$ will
be consistent with uniform based on a Kolmogorov-Smirnov (KS) test.
To take an extreme example, if for a given trial function $F(q)$,
each of the $r_i$ were exactly equal to 0.58, then Equation~(\ref{eqn:rbar})
would be satisfied, but the distribution would not be consistent with uniform
$(p<1\%)$.
Nevertheless, since KS is a relatively weak test, it would be surprising
if a function that satisfied Equation~(\ref{eqn:rbar}) did not meet this
second criterion as well.

We begin by considering power laws $F(q)\propto q^p$.  Applying
Equation~(\ref{eqn:rbar}), we find that
\begin{equation}
p = 1.05^{+0.78}_{-0.68} \qquad \rm (this\ work).
\label{eqn:peval}
\end{equation}

\begin{figure}
\vspace{-6cm}
\plotone{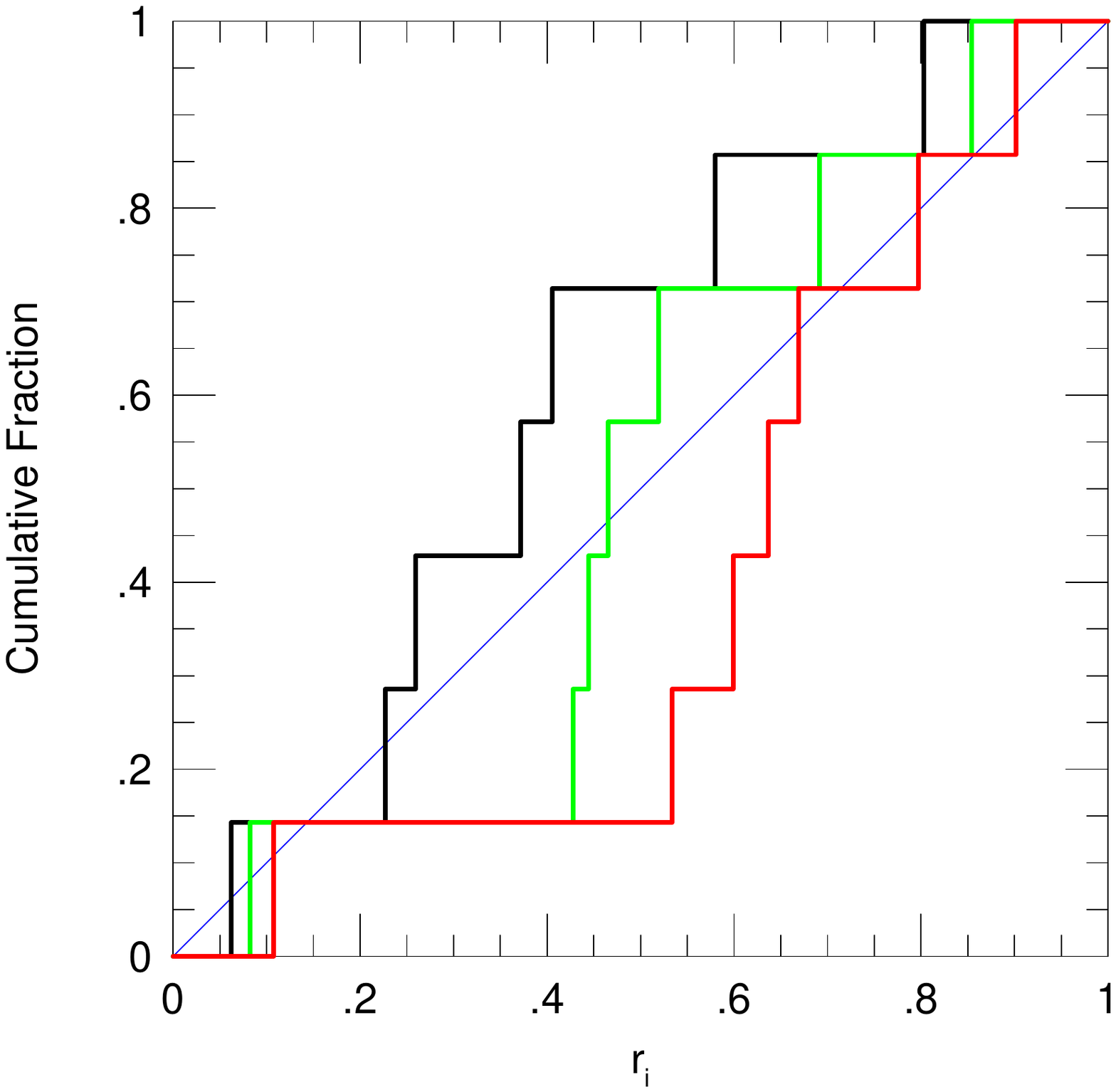}
\caption{Cumulative distribution of the ``$V/V_\max$'' parameter $r_i$ defined
by Equations~(\ref{eqn:vvmax}) and (\ref{eqn:vvmaxspec}) for three power laws
$dN/d\ln q \propto q^p$, where $q$ is the mass ratio and 
$p=1.05$ (green), $p=0.37$ (black), and $p=1.83$ (red).  These represent
the best fit and $1\,\sigma$ lower and upper limits, respectively.  In
all cases, a Kolmogorov-Smirnov test shows that these are consistent with
being drawn from a uniform distribution.  Hence, there is no basis to
reject a power-law for the mass-ratio function from this analysis.
The best fit value confirms a sharp turnover in the mass-ratio function
relative to that found by \citet{ob07368} and \citet{suzuki16} at higher
mass ratios.
}
\label{fig:ks}
\end{figure}

Figure~\ref{fig:ks} shows the cumulative distribution at the best
fit and $1\,\sigma$ limits displayed in Equation~(\ref{eqn:peval}).
These have maximal differences (relative to uniform) of
$D=(0.285,0.309,0.391)$ with corresponding KS $p$-values
$p=(0.53,0.43,0.18)$.  That is, there is no independent information
from the uniformity (or lack of it) that would indicate that
any of these functions is unacceptable at the $1\,\sigma$ level.
Therefore, there is also no basis for rejecting a power-law form for
the distribution in the domain probed by our sample.

\begin{figure}
\vspace{-6cm}
\plotone{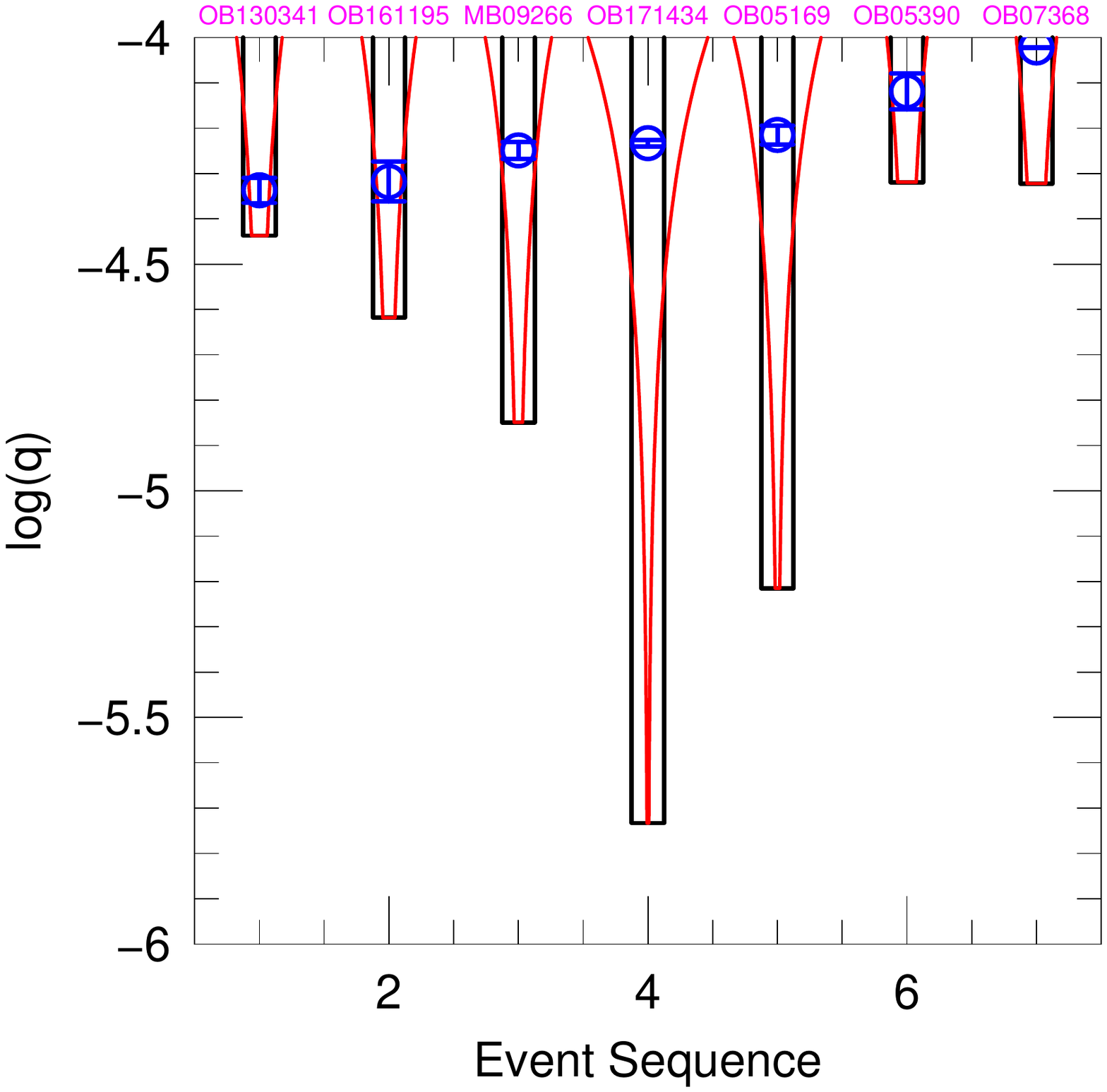}
\caption{Illustration of the ``$V/V_\max$'' method.  The blue circles
show the best fit $q$ from the actual event, while the bottoms of the
black rectangles show the lowest $q^\prime$ that could have been detected.
The red curves show the relative frequency of different mass ratios
according to the best-fit power law, $dN/d\ln q \propto q^{1.05}$.
These can be compared to the relative frequencies that would be expected
from a hypothetical law $dN/d\ln q =$const (black). If the frequency function
is chosen correctly, then on average, half of the red ``volume'' should
be above the blue points.  More generally, the ratio $r_i$ of this ``volume''
to the total ``volume'' should be consistent with being uniformly distributed
over the interval [0,1].
}
\label{fig:vovmax}
\end{figure}

Figure~\ref{fig:vovmax} illustrates the ``$V/V_\max$'' method as well
as the best fit result.  For each event (listed at the top), the
observed mass ratio $q$ is shown by a blue point, while the
lowest $q^\prime$ at which it could have been detected is shown
by the bottom of the rectangular box.  The boxes themselves have
uniform width, which illustrates the relative frequency of $q^\prime$
values for the hypothetical case $dN/d\ln q \propto q^0 = {\rm const}$,
i.e., $p=0$.  The red curves show the relative frequency for the best-fit
case, $p=1.05$.  The parameter $r_i$ is the ratio of the volume ``$V$''
(i.e., area) contained within the red curve above the actual detection,
divided by the total volume ``$V_\max$'' within the red curve.  The
best-fit value ($p=1.05$), illustrated 
in the figure, occurs when $\langle r_i\rangle = 0.5$.

Equation~(\ref{eqn:peval}) is consistent with 
the results of \citet{suzuki16}, who found 
$p=0.6^{+0.5}_{-0.4}$ based on an almost completely independent argument.
Since these arguments are essentially independent, we can combine the two
measurements (weighted according to the quoted errors) to obtain
\begin{equation}
p = 0.73^{+0.42}_{-0.34} \qquad \rm (this\ work\ +\ Suzuki\ et \ al.\ 2016)
\label{eqn:peval2}
\end{equation}

\subsection{$\Delta\chi^2$ at Threshold}
\label{sec:thresh}

In contrast to the present investigation, all previous microlensing studies of
the planet mass or mass-ratio function have used a detection-efficiency 
analysis, which compares the detected planets to a
calculation of the overall planet sensitivity of the sample. The vast
majority of these analyses calculate the planet sensitivity either by
fitting planetary models to the data following \citet{gaudi00}
or by simulating light curves with planets following
\citet{rhie00}. In both cases, the ``detectability'' of a given planet
is assessed relative to a fixed $\Delta\chi^2$ threshold comparing
1L1S models to 2L1S\footnote{The exception is \citet{shvartzvald16},
who used a local $\chi^2$ excess to determine detectability with the
threshold determined individually for each event.}. Early analyses
used $\Delta\chi^2 = 60$ as their detection threshold \citep{albrow01,gaudi02},
but other studies have used thresholds ranging from
$\Delta\chi^2=60$ to $\Delta\chi^2=500$ \citep{ob07368,gould10,
cassan12,suzuki16}. In the analysis of individual events, 
different authors have used different
thresholds depending on the exact event and on whether or not the
planet signal comes from a central caustic or a planetary caustic
(e.g., Section 8 of \citealt{ob07368} and references therein). For example,
\citet{mb10311} argued based on the analysis of MOA-2010-BLG-311 that
$\Delta\chi^2 \sim 80$ is insufficient to claim a secure planet
detection, even though this is above the $\Delta\chi^2$ threshold
used in many studies. They also suggested that this event is
evidence that the threshold for
high-magnification/central-caustic crossing perturbations may be
higher than for planetary caustic perturbations.

Although the $V/V_\max$ analysis presented here was designed to
answer a different question, we can use the results of this analysis
to explore $\Delta\chi^2$ at the threshold of planet
detection.
Table~\ref{tab:chi2} shows
$\Delta\chi^2\equiv\chi^2$(1L1S)$ -\chi^2$(2L1S)
[or $\chi^2$(2L1S)$ -\chi^2$(3L1S) in the case of OGLE-2013-BLG-0341]
at our adopted threshold of ``publishability''
for the eight events considered here (including two models for
OGLE-2017-BLG-0173).  The most striking feature of this table
is that $\Delta\chi^2$ at threshold spans a factor of 100.
As we discuss below, the extreme broadness of this range is partly
due to the heterogeneous character of the sample.  However, even 
after accounting for this heterogeneity, Table~\ref{tab:chi2} strongly
suggests that $\Delta\chi^2$ is a relatively poor proxy for ``publishability''.
Before we begin this review, we emphasize that we regard 
``publishability'' as a more appropriate criterion that ``detectability''
because anomalies can be ``detectable'' and unambiguously real,
yet still be uninterpretable at a level that is scientifically 
interesting.

The four events for which the threshold was $\Delta\chi^2<500$
shed the most light on this question:
OGLE-2005-BLG-169,
OGLE-2016-BLG-1195,
OGLE-2013-BLG-0341, and
OGLE-2005-BLG-390, 
with $\Delta\chi^2= 184$, 257, 309, and 457, respectively.
These are naturally grouped into two pairs:
OGLE-2005-BLG-169 and OGLE-2013-BLG-0341 were both identified from
very short features whose signatures were unambiguously planetary,
while for both OGLE-2016-BLG-1195 and OGLE-2005-BLG-390, the threshold
was set by confusion with 1L2S models.

As discussed in Section~\ref{sec:ob05169}, the actual decision to
publish OGLE-2005-BLG-169 was based on the secure recognition of a 
discontinuity 
in the slope of the light curve.  Such discontinuities cannot be
produced by any microlensing effect other than a planetary (or binary)
companion and would be extremely difficult to generate by instrumental
or weather problems.  We found in Section~\ref{sec:ob05169} that at
the threshold of ``by-eye confidence'' in the reality of this feature,
the mass-ratio error $\sigma(\log q)=0.10$ is relatively close
to our adopted threshold of $\sigma(\log q)<0.15$.  Thus, by two
independent modes of assessment, this event would have been ``barely
publishable'' at the adopted threshold.

For OGLE-2013-BLG-0341, the key signature was an isolated dip, 
which can only be produced by a minor-image perturbation.  Moreover,
such short-lived dips can only be produced by planets.  While the
total $\Delta\chi^2$ of the planet is higher than that of 
OGLE-2005-BLG-169 by 125,
this difference is somewhat deceptive.  Recall from 
Section~\ref{sec:ob130341} that $\Delta\chi^2($binary-caustic$)=66$
came from the perturbation induced by the binary caustic.  However,
this binary-caustic ``confirmation'' played no role in our assessment
of whether the planet would have been published at the adopted threshold.
Hence, if the system had been a simple 2L1S system, or if the trajectory
happened to miss the central caustic, the decision would have been
exactly the same.  In these cases, $\Delta\chi^2=309-66=243$.
Moreover, we have not attempted to assess the exact $\log q^\prime/q$ at which
the event would have ceased to be publishable, mainly because this
level of precision would not contribute significantly to our study,
but also because we do not believe that such precision is feasible.  Hence,
these two values (184 and 243) for these two ``short timescale feature''
events should be regarded as comparable.

In brief, based on this admittedly small sample of two, it appears
feasible to publish ``short timescale feature'' events with 
$\Delta\chi^2\ga 200$.

It is quite notable that the two events whose threshold is set
by confusion with 1L2S models have substantially different $\Delta\chi^2$
in Table~\ref{tab:chi2},
i.e., 257 versus 457.  It is true OGLE-2005-BLG-390 was subjected
to a slightly stronger $\Delta\chi^2$ threshold to reject 1L2S 
($\Delta\chi^2>16$ versus 12), which drove up its threshold of
acceptance from $\log(q'/q)=-0.25$ to $\log(q'/q)=-0.20$.  However, it is
also the case that at the finally adopted threshold for OGLE-2013-BLG-1195,
it met essentially the same threshold (see Section~\ref{sec:ob161195}).
Hence, this is not a major factor in this difference.  Furthermore,
at their respective thresholds, which differ in $\Delta\chi^2$
by a factor 1.8, both events have very similar $\sigma(\log q)\simeq 0.1$.
Hence, based on this very small sample of two ``1L2S-limited'' events,
we already see a very significant difference in $\Delta\chi^2$ at threshold.

The remaining four events have larger $\Delta\chi^2$ between the 1L1S
and 2L1S solutions at the threshold. However, because of their more
complex observation history, interpreting their significance for
understanding $\Delta\chi^2$ thresholds in general would require
significantly more investigation than provided by the $V/V_\max$
analysis. Nevertheless, we review here what is known from the present
analysis.

For OGLE-2008-BLG-0368, the $\Delta\chi^2=701$ appears substantially
larger than the previous four cases, particularly because
it contains a short-duration ``dip'', which we noted above is
a signature of a minor image perturbation that is very difficult to
mimic by other (non-planetary) effects.  Recall from
Section~\ref{sec:ob07368}, however, that the threshold was set by the
availability of follow-up data.  Therefore, while we do not investigate
this issue in detail, we can be confident that the $\Delta\chi^2$
threshold for a similar event with data acquisition that did not
depend on alerts (i.e., without triggered, followup data),
would be significantly smaller.  That is, this
event does not, in itself, provide evidence for a substantially
broader range of $\Delta\chi^2$ threshold than has already been established
above.

The two variants of OGLE-2017-BLG-0173 both have $\Delta\chi^2\sim 1000$,
which is again relatively high compared to the four lowest-$\Delta\chi^2$
events.  This high $\Delta\chi^2$ results directly from our assessment
that a bump with amplitude $\Delta I\la 0.06\,$mag would not have been
noticed in the current mode of KMTNet review of OGLE-discovered events.
However, this bump was very densely covered from two KMTNet
observatories. We hypothesize that a future, systematic,
algorithm-based search for anomalies (rather than the by-eye search)
would likely have discovered such a bump even at substantially lower
amplitude. Although ``potential'' (i.e., future) discoveries are irrelevant to
the $V/V_{\rm max}$ analysis, they are relevant to the question of
establishing an appropriate $\Delta\chi^2$ threshold. The enhanced
sensitivity of such searches would easily bring the threshold for this
event into the $\Delta\chi^2$ range of events like OGLE-2005-BLG-390.
Hence, in the context of a future, machine search for anomalies, the
present analysis provides only an upper limit for the $\Delta\chi^2$
threshold.

The case of OGLE-2017-BLG-1434, with $\Delta\chi^2=4011$, is even more
stark.  Part of the issue here is that for purposes of the current
study, we were forced to consider ``quick look'' data, which were
reduced for only one of the two KMTNet fields, while the
reported $\Delta\chi^2$ value is
based on data from both fields.  Hence, from the standpoint of studying
thresholds, we should really consider that this event has $\Delta\chi^2=2005$.
Even so, this is quite high relative to the first four events that
we examined above.  Recall from Section~\ref{sec:ob171434}, that we
rejected $\log(q'/q)=-1.75$ because, within the context of the visual
reviews by which planets are currently being discovered in KMTNet data,
this would have appeared to have had a single strong outlier, with
a few other deviating points that could easily be taken for noise.
A machine search would easily identify the planetary anomaly
at $\log(q'/q)=-1.75$ in the data shown in Figure~\ref{fig:ob171434},
and this would have then led to publication, even if only data from
one KMTNet field were available.  Thus, as with OGLE-2017-BLG-0173, 
the $\Delta\chi^2$
threshold for a more homogeneous search would be much lower than in
the current study, which provides only an upper limit on the threshold.

Finally, the very high $\Delta\chi^2=20345$ for MOA-2009-BLG-266 is
an order of magnitude larger than for any 
other events considered here.  A major factor is that,
as discussed in Section~\ref{sec:mb09266}, the threshold is set by the 
requirement of an alert, which then greatly increased the $\Delta\chi^2$
at the threshold.  Further investigation and assessment of the influence
of followup data on the $\Delta\chi^2$ threshold in this potentially interesting
case is outside the scope of the present study.

The $\Delta\chi^2$
values from Table~\ref{tab:chi2} provide an interesting window
into the role of $\Delta\chi^2$ as a proxy for publishability.
However, because the current study is founded on an inhomogeneous
sample of planetary events, this review of these $\Delta\chi^2$
cannot be regarded as a definitive investigation of this question.  In our
view, the review provides evidence that the $\Delta\chi^2$
threshold for homogeneous samples is likely to vary at the factor 2
level but a more focused study would be required to confirm this.

\subsection{Host Mass of Low-$q$ Planets}
\label{sec:hostmass}

\begin{figure}
\vspace{-6cm}
\plotone{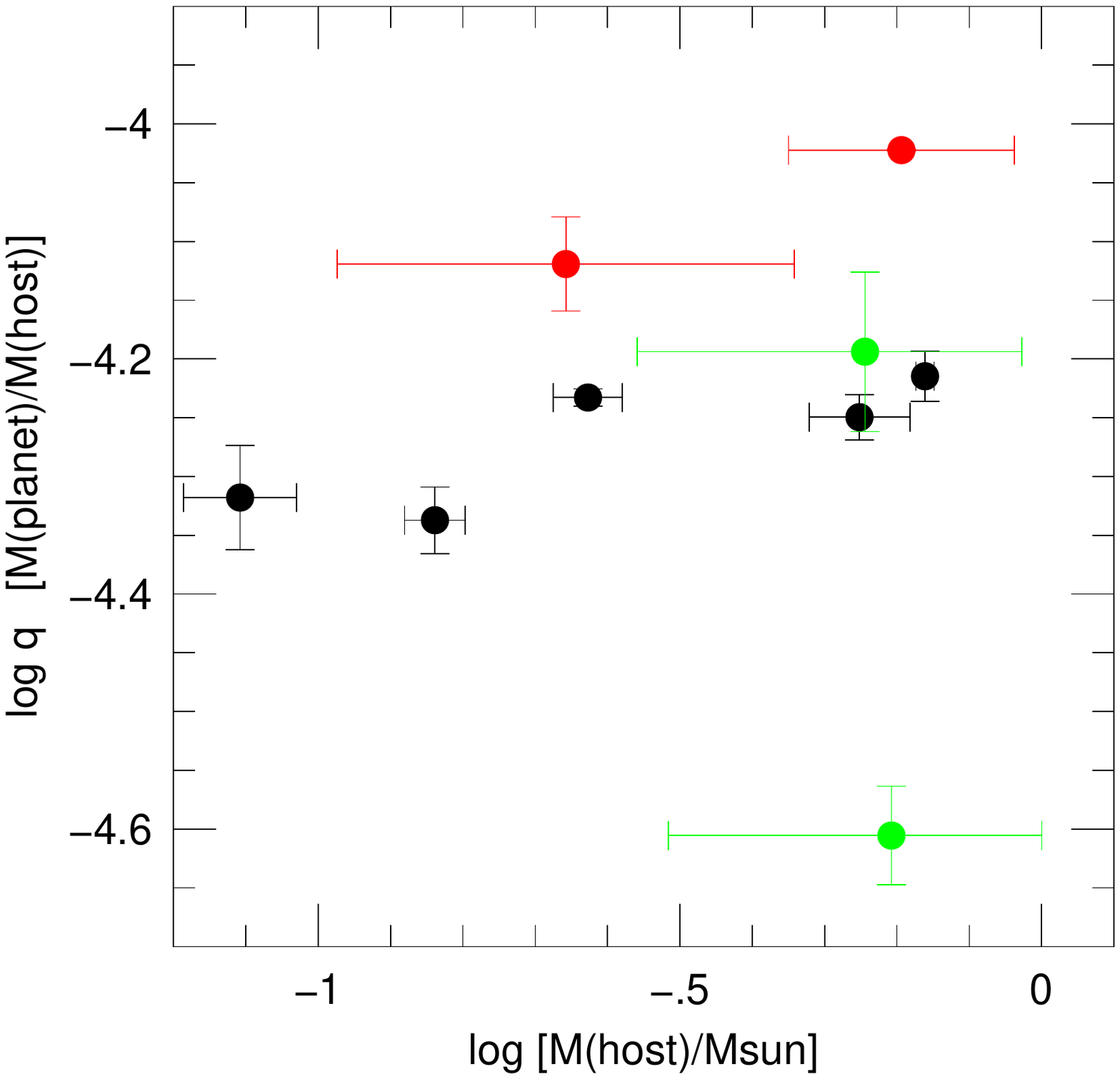}
\caption{Mass ratio $q$ vs.\ host mass $M$ for eight microlensing
planets with best-fit mass ratios $q<10^{-4}$.  The black points
have well-measured host masses, either from microlens parallax or
direct imaging (left to right, 
OGLE-2016-BLG-1195,
OGLE-2013-BLG-0341,
OGLE-2017-BLG-1434,
MOA-2009-BLG-266, and
OGLE-2005-BLG-169).  
The red points have host masses derived from Bayesian estimates 
(OGLE-2005-BLG-390 and OGLE-2007-BLG-368).  The green points show
two different degenerate models for OGLE-2017-BLG-0173, which is
not included in our statistical sample.  The range of host masses is
one dex while the range of mass ratios is one octave, suggesting
that planet-host mass ratio is a better defined function than planet
mass.
}
\label{fig:mq}
\end{figure}

Figure~\ref{fig:mq} shows the planet-host mass ratio $q$ versus host
mass$M$ for the eight planets with low mass ratios: $q<10^{-4}$.  Microlenses
with well measured host masses (either from parallax or direct imaging)
are shown in black, while those with Bayesian estimates are shown in red.
The two blue points show the ambiguous $q$ determination for OGLE-2017-BLG-0173,
which is excluded from the present study because of this ambiguity, but
is displayed here for completeness\footnote{The host masses for the two
solutions of OGLE-2017-BLG-0173 are corrected relative to those given by
\citet{ob170173}, which were impacted by a bug in the Bayesian code.
In particular, the median of the correct estimates of the mass are about
1.5 times higher than the original.  These correction have no impact on the 
present study, in part because it deals with mass ratios (which are unaffected)
rather than masses, but also because OGLE-2017-BLG-0173 is excluded from
the study.  However, it does impact the display in Figure~\ref{fig:mq}.
For completeness, we report here the corrected Bayesian estimates for
models (A,B,C): 
$M_{\rm host} = 
(0.57_{-0.29}^{+0.38},
0.62_{-0.32}^{+0.38},
0.62_{-0.31}^{+0.37})M_\odot$,
$M_{\rm planet} = 
(12_{-7}^{+11},
5.1_{-2.8}^{+5.1},
14_{-8}^{+11})M_\oplus$,
$D_{\rm L} = 
(5.9_{-1.5}^{+1.0},
5.5_{-1.5}^{+1.1},
5.6_{-1.5}^{+1.0})\kpc$,
$a_\perp = 
(4.3_{-1.3}^{+1.2},
4.6_{-1.5}^{+1.3},
4.4_{-1.4}^{+1.2})\au$.
}.

The main implication of this figure is that the range in $\log M$ is
about three times larger than in $\log q$.  The broad range in $\log M$
is not the result of measurement errors: it would remain the same even
if the two red points were excluded.  In this section, we have shown
(see Figure~\ref{fig:vovmax}) that the restricted range in $\log q$ 
is not a selection effect.  That is, many (though not all) of these planets
could have been detected at much lower $q$.  Figure~\ref{fig:mq} suggests,
although it certainly does not prove, that the planet mass-ratio function
is a better framework for understanding planet masses than the planet
mass function.  That is, it suggests that the turnover first pointed
out by \citet{suzuki16}, and confirmed in this section, is a turnover
in planet mass ratio rather than in planet mass.

In this context, it is interesting to note that in a study
carried out contemporaneously with the present one,
\citet{kepq} found that the mass-ratio function
derived from {\it Kepler} planets with periods $P<100\,$days
is independent of host mass for hosts $M<1\,M_\odot$.  They found
a broken power law, with a slope at the low-mass end that is
consistent with those derived here and by \citet{suzuki16}, but with
the break roughly a decade below the one found by \citet{suzuki16}.
In brief, there is some evidence from the present study that
planet-host mass ratio governs planet formation outside the snow line
and stronger evidence from \citet{kepq} that this is the case inside 
the snow line, even though planet formation peaks at very different
mass ratios in these two zones.

Finally, we note that one of the solutions for OGLE-2017-BLG-0173
(blue points) has similar $q$ to those of the seven other points,
while the other is separated from the entire group by almost 0.3 dex.
While no strong conclusion can be drawn from this, it suggests that
the higher mass-ratio solution is correct.  Unfortunately, as
discussed by \citet{ob170173}, it is quite unlikely that this degeneracy
will be resolved by future followup observations.

\section{Conclusions}
\label{sec:conclude}

With a planet-host mass ratio $q=5.8\times 10^{-5}$ and planet mass 
$m_p= 4.4\,M_\oplus$, OGLE-2017-BLG-1434Lb facilitates a new probe
of cold, low-mass planets.  It is the eighth microlensing
planet with $q<10^{-4}$.  Combining seven of these detections, and applying
a new ``$V/V_\max$'' argument, we have shown that the planet-host
mass-ratio function turns over at low mass.  That is, it rises sharply toward
lower mass for $q\ga 2\times 10^{-4}$ (power law $n\sim -1$) but then
falls just as sharply toward lower mass for $q< 10^{-4}$ 
(power law $p\sim +1$).

\acknowledgments 
Work by WZ, YKJ, and AG were supported by AST-1516842 from the US NSF.
WZ, IGS, and AG were supported by JPL grant 1500811.  
Work by C.H. was supported by the grant (2017R1A4A1015178) of
National Research Foundation of Korea.
This research has made use of the KMTNet system operated by the Korea
Astronomy and Space Science Institute (KASI) and the data were obtained at
three host sites of CTIO in Chile, SAAO in South Africa, and SSO in
Australia.
The OGLE project has received funding from the National Science Centre, 
Poland, grant MAESTRO 2014/14/A/ST9/00121 to AU.
Work by YS was supported by an appointment to the NASA Postdoctoral Program at the Jet Propulsion Laboratory,
California Institute of Technology, administered by Universities Space Research Association
through a contract with NASA.

\begin{deluxetable}{lcccccc}
\tablecolumns{7} \rotate
\tablewidth{0pc}\tablecaption{\textsc{Best-fit solutions}}
\tablehead{ \colhead{ } & \colhead{ }&
\multicolumn{2}{c}{Parallax models}& \colhead{ }&\multicolumn{2}{c}{Parallax+Orbital motion models}\\
\cline{3-4} \cline{6-7} \colhead{Parameters } & \colhead{Standard}&
\colhead{$u_0>1$}&\colhead{$u_0<1$}& \colhead{
}&\colhead{$u_0>1$}&\colhead{$u_0<1$}} \startdata
  $\chi^2/\rm{dof}$               &21418.644/18659       &18658.511/18657       &18662.277/18657        & &18654.143/18655       &18658.155/18655       \\
  $t_0$ $(\rm{HJD}^{\prime})$     &7984.935 $\pm$ 0.004  &7984.979 $\pm$ 0.004  &7984.978 $\pm$ 0.004   & &7984.978 $\pm$ 0.004  &7984.977 $\pm$ 0.004  \\
  $u_0$                           &0.037 $\pm$ 0.001     &0.044 $\pm$ 0.001     &-0.044 $\pm$ 0.001     & &0.043 $\pm$ 0.001     &-0.043 $\pm$ 0.001    \\
  $t_{\rm E}$ $(\rm{days})$       &72.856 $\pm$ 0.907    &61.421 $\pm$ 0.692    &62.981 $\pm$ 0.788     & &62.957 $\pm$ 0.863    &64.255 $\pm$ 1.006    \\
  $s$                             &0.980 $\pm$ 0.0003    &0.978 $\pm$ 0.0003    &0.978 $\pm$ 0.0003     & &0.979 $\pm$ 0.0004    &0.979 $\pm$ 0.0003    \\
  $q$ $(10^{-5})$                 &4.938 $\pm$ 0.057     &5.866 $\pm$ 0.063     &5.571 $\pm$ 0.077      & &5.722 $\pm$ 0.145     &5.607 $\pm$ 0.152     \\
  $\alpha$ $(\rm{rad})$           &4.535 $\pm$ 0.001     &4.552 $\pm$ 0.001     &-4.556 $\pm$ 0.002     & &4.551 $\pm$ 0.002     &-4.553 $\pm$ 0.002    \\
  $\rho$ $(10^{-4})$              &4.019 $\pm$ 0.060     &4.815 $\pm$ 0.068     &4.679 $\pm$ 0.072      & &4.692 $\pm$ 0.093     &4.643 $\pm$ 0.099     \\
  $\pi_{\rm{E},\it{N}}$           &-                     &-0.491 $\pm$ 0.079    &-0.508 $\pm$ 0.083     & &-0.586 $\pm$ 0.081    &-0.562 $\pm$ 0.081    \\
  $\pi_{\rm{E},\it{E}}$           &-                     &0.471 $\pm$ 0.013     &0.475 $\pm$ 0.013      & &0.472 $\pm$ 0.013     &0.471 $\pm$ 0.013     \\
  $ds/dt$ $(\rm{yr}^{-1})$        &-                     &-                     &-                      & &0.069 $\pm$ 0.044     &0.090 $\pm$ 0.044     \\
  $d\alpha/dt$ $(\rm{yr}^{-1})$   &-                     &-                     &-                      & &-0.218 $\pm$ 1.432    &-1.543 $\pm$ 1.459    \\
  $f_S$                           &0.139 $\pm$ 0.002     &0.167 $\pm$ 0.002     &0.166 $\pm$ 0.002      & &0.162 $\pm$ 0.003     &0.163 $\pm$ 0.003     \\
  $f_B$                           &0.193 $\pm$ 0.002     &0.165 $\pm$ 0.002     &0.166 $\pm$ 0.002      & &0.170 $\pm$ 0.003     &0.169 $\pm$ 0.003     \\
\enddata
\tablecomments{In the ``Parallax+Orbital'' models, the parameter
$\beta\equiv {\rm (KE/PE)}_\perp$ is restricted to $\beta<0.7$.  See text.}
\label{tab:ulens}
\end{deluxetable}

\begin{deluxetable}{lccccccc}
\tablecolumns{8}
\tablewidth{0pc}\tablecaption{\textsc{Characteristics of
Earth/Super-Earth Events}}
\tablehead{\colhead{Event}&\colhead{$q(10^{-5})$} &\colhead{$s$}
&\colhead{$M_{\rm p}/M_\earth$} &\colhead{$M_{\rm h}/M_\sun$}
&\colhead{$D_{\rm L}$/kpc} &\colhead{$a_\bot$/au}
&\colhead{$a_\bot$/$a_{\rm snow}$}} \startdata
 OB161195      &4.81  &0.99  &1.43  &0.078  &3.91  &1.16  &5.5 \\
 OB130341      &4.60  &1.00  &2.00  &0.145  &1.16  &0.88  &3.0 \\
 OB161434      &5.72  &0.98  &4.48  &0.232  &0.87  &1.18  &1.9 \\
 \enddata
\label{tab:lowm}
\end{deluxetable}

\begin{deluxetable}{lrrrr}
\tablecolumns{5} \tablewidth{0pc}\tablecaption{\textsc{$\Delta
\chi^2$}} \tablehead{\colhead{Event}&\colhead{log($q'/q$)}
&\colhead{$\chi^2$(2L1S)} &\colhead{$\chi^2$(1L1S)}
&\colhead{$\Delta\chi^2$}} \startdata
 OB05169       &-1.0  &506    &690    &184   \\
 OB161195      &-0.3  &12415  &12672  &257   \\
 OB130341      &-0.1  &8874   &9183   &309   \\
 OB05390       &-0.2  &551    &1008   &457   \\
 OB07368       &-0.3  &2625   &3326   &701   \\
 OB170173 (A)  &-0.7  &7435   &8389   &954   \\
 OB170173 (B)  &-0.7  &7432   &8582   &1150   \\
 MB09266       &-0.6  &4153   &24497  &20344   \\
 OB171434      &-1.5  &18236  &22247  &4011   \\
\label{tab:chi2}
\enddata
\end{deluxetable}

\end{document}